\documentclass[aps,prd,preprintnumbers,superscriptaddress,nofootinbib,notitlepage,floatfix,10pt]{revtex4-2}
\usepackage[pdftex]{graphicx}
\usepackage{bm,latexsym,amsmath,amssymb,amsfonts,mathrsfs}
\usepackage{color}
\allowdisplaybreaks[1]
\usepackage[pdftex,colorlinks=true,linkcolor=blue,citecolor=cyan,backref=page]{hyperref}
\usepackage{amsthm}
\theoremstyle{plain}
\newtheorem*{prop*}{Proposition}
\usepackage{enumitem}
\newcommand*{\D}{\mathrm{d}}

\begin{document}
\title{Healthy scalar-tensor theories with third-order derivatives:
Generalized disformal Horndeski and beyond}
%
\author{Masaki~Michiwaki}
\email[Email: ]{michiwaki@rikkyo.ac.jp}
\affiliation{Department of Physics, Rikkyo University, Toshima, Tokyo 171-8501, Japan}
\author{Tsutomu~Kobayashi}
\email[Email: ]{tsutomu@rikkyo.ac.jp}
\affiliation{Department of Physics, Rikkyo University, Toshima, Tokyo 171-8501, Japan}
%
\begin{abstract}
We systematically construct a broad class of ghost-free scalar-tensor theories whose 
Lagrangian includes up to third-order derivatives 
of the scalar field. Using a spatially 
covariant action written in terms of the ADM 
variables, we impose degeneracy and consistency 
conditions that remove the additional Ostrogradsky 
degree of freedom and leave one scalar and two 
tensor degrees of freedom within the regular branch 
considered in this paper. 
The resulting class extends the generalized 
disformal Horndeski and U-DHOST families.
\end{abstract}
\preprint{RUP-26-1}
    \maketitle

\section{Introduction and summary}\label{sec:intro}

Scalar-tensor theories are modified theories of gravity in which one introduces one or more 
scalar degrees of freedom in addition to the dynamical metric. 
Phenomenologically, scalar-tensor theories provide models for cosmological inflation and dark energy,
and form a theoretical framework for black holes and other astrophysical objects beyond general relativity.
Extending the theory space of scalar-tensor gravity and clarifying its structure are important for exploring the fundamental nature of gravity.

Efforts have been made to extend the scalar-tensor theory space since the rediscovery of the Horndeski theory, the most general scalar-tensor theory having second-order equations of motion~\cite{Horndeski:1974wa, Deffayet:2011gz, Kobayashi:2011nu} (see Ref.~\cite{Kobayashi:2019hrl} for a review).
One way to enlarge 
the theory space is to include terms containing higher-order derivatives of the scalar field. 
In general, if the equations of motion contain higher-order time derivatives, 
the theory propagates additional unstable degrees of freedom known as Ostrogradsky ghosts~\cite{Ostrogradsky:1850fid, Woodard:2015zca}. If, however, the kinetic matrix of the Lagrangian is degenerate, constraint equations arise, by which one can eliminate the extra degrees of freedom.

To construct higher-derivative scalar-tensor theories without Ostrogradsky ghosts, several ideas have been proposed so far.
The first idea is to write down a covariant Lagrangian and impose degeneracy conditions.
This leads to ghost-free theories, known as degenerate higher-order scalar-tensor (DHOST) theories~\cite{Langlois:2015cwa, Crisostomi:2016czh, BenAchour:2016fzp}.
The well-studied examples include quadratic/cubic DHOST theories having Lagrangians constructed out of terms quadratic/cubic in the second-order derivatives of the scalar field. 
In general, for Lagrangians involving derivatives higher than second order, 
one needs extra conditions in addition to the degeneracy conditions to remove all Ostrogradsky ghosts~\cite{Motohashi:2014opa,Motohashi:2016ftl,Motohashi:2017eya, Motohashi:2018pxg}.
The second idea is to relax the spacetime diffeomorphism symmetry and consider theories
that are invariant only under the spatial diffeomorphism. 
Such theories are called spatially covariant gravity~\cite{Gao:2014soa, Gao:2014fra, Gao:2018znj}, and the basic idea can be traced back to Refs.~\cite{Arkani-Hamed:2003pdi, Cheung:2007st, Horava:2009uw}.
In a spatially covariant theory, the presence of a scalar field is not manifest in the action.
Nevertheless, one can build theories with a healthy\footnote{Here and below, ``healthy" or ``ghost-free" refer only to the absence of the
additional Ostrogradsky degree of freedom, as established by the
constraint analysis. It does not imply stability of the remaining
degrees of freedom, hyperbolicity, or well-posedness of the
initial-value problem.} scalar degree of freedom by performing a Hamiltonian analysis to count the dynamical degrees of freedom.
Another way of seeing this is to regard spatially covariant action as a gauge-fixed version of some fully covariant scalar-tensor theory.
Starting from a spatially covariant theory, one can restore the time diffeomorphism invariance via the St\"{u}ckelberg trick to obtain the corresponding fully covariant scalar-tensor theory.
Since it suffices for the action to be degenerate in a particular gauge (the unitary gauge), the spatially covariant framework can cover broader classes than DHOST theories, resulting in the U-DHOST family of theories~\cite{DeFelice:2018ewo, DeFelice:2021hps}.
The third idea is to apply an invertible field redefinition to some known scalar-tensor theory.
This way of deriving new theories is less systematic but straightforward once such an appropriate field transformation is found.
A well-known example is the disformal transformation, i.e., the most general metric transformation involving the first derivatives of the scalar field~\cite{Bekenstein:1992pj, Bruneton:2007si, Bettoni:2013diz, Zumalacarregui:2013pma, Domenech:2015tca, BenAchour:2016cay, Takahashi:2017zgr}:
\begin{align}
    g_{\mu\nu}\to g_{\mu\nu}'=\Omega(\phi,X)g_{\mu\nu}+\Gamma(\phi,X)\nabla_\mu\phi 
    \nabla_\nu\phi,\label{def:disformalT}
\end{align}
where $\phi$ is a scalar field and
$\Omega$ and $\Gamma$ are functions of $\phi$ and $X:=g^{\mu\nu}\nabla_\mu\phi\nabla_\nu\phi$.
If $\Omega$ and $\Gamma$ satisfy certain conditions, the transformation is invertible.
Using disformal transformations, one can generate a subclass of quadratic/cubic DHOST theories from the Horndeski theory~\cite{Crisostomi:2016czh, BenAchour:2016fzp}.
The subclass thus generated coincides with the cosmologically viable subset of DHOST theories, while theories disformally disconnected from the Horndeski theory exhibit pathologies in cosmological solutions~\cite{deRham:2016wji,Langlois:2017mxy}.
Note that, in the absence of matter, an invertible field transformation yields an equivalent theory with the same number of dynamical degrees of freedom~\cite{Takahashi:2017zgr}.
Once matter fields are introduced, however, the two theories related via the transformation are no longer equivalent because of different couplings to matter when compared in the same frame.
This fact not only gives rise to interesting phenomenology~\cite{Kobayashi:2014ida,Koyama:2015oma,Saito:2015fza,Crisostomi:2017lbg,Langlois:2017dyl,Dima:2017pwp,Hirano:2019scf,Crisostomi:2019yfo} but also prohibits the addition of a certain kind of matter fields to DHOST theories~\cite{Deffayet:2020ypa,Garcia-Saenz:2021acj,Domenech:2025gao}.

The key to generating new higher-order scalar-tensor theories lies in finding a new invertible metric transformation involving higher derivatives of the scalar field.
In recent years, the disformal transformation~\eqref{def:disformalT} has been generalized to allow for the dependence on 
higher-order derivatives of the scalar field, leading to the formulation of generalized 
disformal Horndeski (GDH) theories generated from the Horndeski theory~\cite{Takahashi:2021ttd, Takahashi:2022mew}.
(See also Ref.~\cite{Babichev:2024eoh} for a generalization of the disformal transformation.)
Specifically, the transformation is now allowed to depend on $\nabla_\mu X$ in a particular way.
The GDH family of theories provides concrete examples of ghost-free theories whose action contains third-order derivatives of the scalar field, and thus can be regarded as an extension of quadratic/cubic DHOST theories.
It has been shown that the GDH family includes a cosmologically viable subclass and allows for consistent matter coupling~\cite{Takahashi:2022mew}.

However, as noted above, theories generated via field redefinitions form only a subset of those obtained through a systematic construction.
It is therefore natural to expect that there exist ghost-free scalar-tensor theories with third-order derivatives beyond GDH theories.
In this paper, we systematically construct ghost-free scalar-tensor theories with third-order derivatives from a spatially covariant action (a unitary gauge action) and thereby extend GDH theories.
Once a spatially covariant action is obtained, it is straightforward to recover full general covariance by means of the St\"{u}ckelberg trick.
Although quadratic/cubic DHOST theories disformally disconnected from Horndeski are pathological in a cosmological setup, the same would not necessarily be true for ``beyond GDH'' theories that cannot be related to the Horndeski theory through generalized disformal (GD) transformations.
It is therefore important to study the structure of the broader theory space than the GDH family of theories.

The purpose of this paper is to construct a broad class of scalar-tensor theories with third-order derivatives beyond the GDH family on a generic, regular branch, rather than to establish an exhaustive classification of all exceptional or singular branches. Accordingly, the term ``general'' below refers to generality within the ansatz and the branch specified in this paper.

Since the rest of the paper is rather technical, we outline here the main results.
To extend the GDH theories systematically without relying on invertible field redefinitions,
we start with a sufficiently general action that covers a broader class of theories than the GDH action,
and apply the results of the Hamiltonian analysis~\cite{Gao:2018znj} to impose the conditions that
enforce the theories to retain one scalar and two tensor degrees of freedom.
For this purpose, we work in the unitary gauge action expressed in terms of the lapse function $N$, the spatial metric $\gamma_{ij}$ on constant-scalar-field hypersurfaces, their respective ``velocities'' $V$ and $K_{ij}$ (the extrinsic curvature), and the spatial covariant derivative $D_i$. 
Our analysis results in the action of the form
\begin{align*}
    S&=\int \D t\D^3x \sqrt{\gamma}N\mathcal{L}(t,N,\gamma_{ij}, V, K_{ij},D_i),
    \\ 
    \mathcal{L}&=\mathcal{K}^{ij}Q_{ij}+
    \frac{1}{2}\mathcal{K}^{ij,kl}Q_{ij}Q_{kl}
    +\mathcal{H}_1^{ij,kl}Q_{ij}D_ka_{l}
    +\mathcal{L}_0,
\end{align*}
where
\begin{align*}
    Q_{ij}&=K_{ij}+\mathcal{U}_{ij}V+\frac{1}{2}\mathcal{V}_{ij}^kD_kV,
    \\ 
    a_i&=D_i\ln N,
\end{align*}
$\mathcal{K}^{ij}$, $\mathcal{K}^{ij,kl}$, and $\mathcal{H}_1^{ij,kl}$ are general tensors built out of $\gamma^{ij}$ and $a^i$, with in total 13 arbitrary functions of $t$, $N$, and $Z=a_ia^i$.
(Explicitly, they are given by Eqs.~\eqref{def:calKij},~\eqref{def:calKijkl}, and~\eqref{def:cal H1}.)
The tensors $\mathcal{U}_{ij}$ and $\mathcal{V}_{ij}^k$ must be of particular forms characterized by two free functions $W_1(t,N,Z)$ and $W_2(t,N,Z)$.
The Lagrangian $\mathcal{L}_0$ contains only terms without time derivatives, such as the three-dimensional Ricci scalar, $R^{(3)}$, multiplied by a function of $t$, $N$, and $Z$.
The above action contains the GDH theory as a subset.
Specifically, choosing particular forms of $\mathcal{K}^{ij,kl}$, $\mathcal{K}^{ij}$, and $\mathcal{H}_1^{ij,kl}$, one can reproduce the GDH theory.

Generalized disformal transformations of the metric can be expressed as
\begin{align*}
    N\to N'=F_0(t,N),
    \qquad 
    N^i\to {N^i}'=N^i+\frac{F_2(t,N,Z)}{F_1+ZF_3}a^i,
    \qquad 
    \gamma_{ij}\to \gamma_{ij}'=F_1(t,N,Z)\gamma_{ij}+F_3(t,N,Z)a_ia_j,
\end{align*}
where $N^i$ is the shift vector.
The form of the above Lagrangian is stable under GD transformations.
Furthermore, by choosing $F_1$ and $F_3$ appropriately, one can move to the frame in which
$\mathcal{U}_{ij}=\mathcal{V}_{ij}^k=0$.

The rest of the paper is organized as follows.
In the next section, we introduce the building blocks 
and present the unitary gauge action that extends 
the GDH family of theories. In Sec.~\ref{sec:degeneracy}, 
we impose the degeneracy conditions that are necessary to 
eliminate an extra Ostrogradsky mode.
In Sec.~\ref{sec:consistency}, we derive the consistency 
conditions, which further restrict the allowed form of 
the Lagrangian. In Sec.~\ref{sec:GD}, we investigate the transformation properties
of our theories under GD transformations, and describe their 
relation to GDH and U-DHOST theories.
Finally, Sec.~\ref{sec:discussion} is devoted to discussion.

\section{Construction of the action}

\subsection{Building blocks}\label{subsec:bb}

The first step is to introduce the fundamental ingredients 
that serve as the building blocks of our action.
In our setup, we consider a four-dimensional spacetime, which is foliated 
by a one-parameter family of spacelike hypersurfaces. 
The foliation parameter is identified with a scalar field $\phi$.
Equivalently, each leaf is a constant-$\phi$ hypersurface. 
The building blocks of our action are constructed 
from the four-dimensional metric $g_{\mu\nu}$ and $\phi_\mu:=\nabla_\mu \phi$.
To write the action, it is convenient to introduce
the unit timelike normal to constant-$\phi$ hypersurfaces,
\begin{align}
    n_\mu:=
    -\mathcal{N}\phi_\mu,
\end{align}
where 
$\mathcal{N}:=1/\sqrt{-X}$ with
$X:=g^{\mu\nu}\phi_\mu\phi_\nu=\phi_\mu\phi^\mu$.
Note that $\pounds_n\phi = \mathcal{N}^{-1}$,
where $\pounds_n$ denotes the Lie derivative along $n_\mu$.
The basic geometric quantities we use are
the extrinsic curvature $K_{\mu\nu}$ on constant-$\phi$ hypersurfaces,
the velocity of $\mathcal{N}$,
and the acceleration vector 
$a_\mu$, given respectively by
\begin{align}
    K_{\mu\nu}&:=\mathcal{P}_\mu^\lambda\mathcal{P}_\nu^\rho
    \nabla_\lambda n_\rho
    =\frac{1}{2} \pounds_n \mathcal{P}_{\mu\nu}
    ,
    \\ 
    V&:=
    -\frac{\phi^\mu\phi^\nu\phi_{\mu\nu}}{X^2}
    =\pounds_n\mathcal{N}
    ,
    \\
    a_\mu&:=n^\nu\nabla_\nu n_\mu
    = \mathcal{P}_\mu^\nu\nabla_\nu \ln \mathcal{N}, 
\end{align}
where we defined the projection tensor
(or the induced metric on constant-$\phi$ hypersurfaces)
$\mathcal{P}_\mu^\nu:=\delta_\mu^\nu+n_\mu n^\nu$
and $\phi_{\mu\nu}:=\nabla_\mu\nabla_\nu\phi$. 
These quantities contain second derivatives of $\phi$.
Conversely, the second derivative of $\phi$ can be expressed using these quantities as
\begin{align}
    \phi_{\mu\nu}=
    \frac{1}{\mathcal{N}}\left(n_\mu a_\nu+n_\nu a_\mu\right)-\frac{V}{\mathcal{N}^2}n_\mu n_\nu
    -\frac{1}{\mathcal{N}}K_{\mu\nu}.
\end{align}

Using the above geometric quantities rather than $\phi_\mu$ and its derivatives,
the Lagrangian for the quadratic DHOST theories~\cite{Langlois:2015cwa} is written as
\begin{align}
    \mathcal{L}_{\text{qDHOST}}&=A_0+A_1K+A_2V+f\mathcal{R}
    +A_3K_{\mu\nu}K^{\mu\nu}+A_4K^2+A_5V^2+A_6 KV+A_7Z,
\end{align}
where $\mathcal{R}$ is the Ricci scalar, $K:=K_\mu^\mu$, and $Z:=a_\mu a^\mu$.
The latter five terms are obtained by contracting the four indices of 
$\phi_{\mu\nu}\phi_{\lambda\rho}$ with $g^{\mu\nu}$ and $\phi^\mu$ in different ways.
The coefficients $A_i$ and $f$ are functions of $\phi$ and $\mathcal{N}$
(or, equivalently, $X$),
and some of them are subject to certain relations called the degeneracy conditions,
so that there are one scalar and two tensor propagating degrees of freedom.

To extend the quadratic DHOST theories,
let us now introduce geometric quantities
corresponding to the third derivatives of $\phi$, $\nabla_\mu\phi_{\nu\lambda}$,
which can be obtained by differentiating $K_{\mu\nu}$, $V$, and $a_\mu$.
Among various possibilities of such quantities, we only consider the following two,
\begin{align}
    V_\mu:=\mathcal{P}_\mu^\nu\nabla_\nu V, \qquad 
    a_{\mu\nu}:=\mathcal{P}_\mu^\lambda\mathcal{P}_\nu^\rho
    \nabla_\lambda a_\rho \,(=a_{\nu\mu}), \label{eq:third_derivatives}
\end{align}
i.e., derivatives of $V$ and $a_\mu$ intrinsic to constant-$\phi$ hypersurfaces.
We assume that third derivatives of the scalar field appear in the action only through
$V_\mu$ and $a_{\mu\nu}$.
The guiding principles behind the construction of these quantities are as follows.
First, we exclude terms involving any further derivatives acting on the induced metric $\mathcal{P}_{\mu\nu}$.
Second, we exclude terms in which derivatives normal to constant-$\phi$ hypersurfaces act on $\mathcal{N}$ more than once.
One might then also include $n^\nu\nabla_\nu a_\mu$
(or $\pounds_na_\mu$).
However, since
$n^\nu\nabla_\nu a_\mu =\mathcal{N}^{-1}V_\mu-a^\nu K_{\mu\nu}+Z n_\mu$,
it is redundant.

We consider the action containing terms up to quadratic in
$K_{\mu\nu}$, $V$, $V_\mu$, and $a_{\mu\nu}$.
Each linear term can be written schematically as
\begin{align}
    A(\phi,\mathcal{N},Z)\cdot C \cdot T,
\end{align}
where $T$ is one of $K_{\mu\nu}$, $V$, $V_\mu$, and $a_{\mu\nu}$,
$C$ is a tensor built out of $g^{\mu\nu}$ and $a^\mu$,
and $A$ is a function of $\phi$, $\mathcal{N}$, and $Z$.
We do not need to include $n^\mu$ in $C$ because we have
$n^\mu K_{\mu\nu}=0$, $n^\mu V_\mu =0$, and $n^\mu a_{\mu\nu}=0$.
Specifically, we have
\begin{align}
  K, \qquad a^\mu a^\nu K_{\mu\nu}, 
  \qquad V, \qquad a^\mu V_\mu, \qquad a^\mu_\mu,
  \qquad a^\mu a^\nu a_{\mu\nu}, \label{L.bb}
\end{align}
multiplied by functions of $\phi$, $\mathcal{N}$, and $Z$.
Similarly, each quadratic term can be written schematically as
\begin{align}
    A(\phi,\mathcal{N},Z) \cdot C\cdot T_1\cdot T_2,
\end{align}
where each of $T_1$ and $T_2$ is one of $K_{\mu\nu}$, $V$, $V_\mu$, and $a_{\mu\nu}$,
$C$ is a tensor built out of $g^{\mu\nu}$ and $a^\mu$,
and $A$ is a function of $\phi$, $\mathcal{N}$, and $Z$.
Specifically, we have
 \begin{align}    
   &K_{\mu\nu}K^{\mu\nu},\qquad 
   K^2,\qquad 
   Ka^\mu a^\nu K_{\mu\nu}, \qquad
  a^\mu K_{\mu\nu}a^\lambda K_\lambda^\nu,\qquad
  \left(a^\mu a^\nu K_{\mu\nu}\right)^2, \nonumber\\
  &V^2 ,\qquad V_\mu V^\mu ,\qquad 
 \left(a^\mu V_\mu\right)^2,\qquad Va^\mu V_\mu, \nonumber\\
 &KV, \qquad 
a^\mu a^\nu K_{\mu\nu}V,\qquad 
 a^\mu K_{\mu\nu}V^\nu, \qquad 
 Ka^\mu V_\mu, \qquad 
  a^\mu a^\nu a^\lambda K_{\mu\nu}V_\lambda,\nonumber\\
 &K_{\mu\nu}a^{\mu\nu},\qquad
 Ka^\mu_\mu, \qquad
 a^\mu a^\nu K_{\mu\nu}a^\lambda_\lambda,\qquad 
 Ka^\mu a^\nu a_{\mu\nu},\qquad
a^\mu a^\nu K_{\mu\nu}a^\lambda a^\rho a_{\lambda\rho},\qquad
 a^\mu K_{\mu\nu}a^\lambda a^\nu_\lambda,\nonumber \\
   &Va^\mu_\mu, \qquad
   Va^\mu a^\nu a_{\mu\nu}, \qquad
   a^\mu_\mu a^\nu V_\nu,\qquad
   a^\mu a_{\mu\nu}V^\nu,  \qquad
   a^\mu a^\nu a^\lambda a_{\mu\nu}V_\lambda,
   \nonumber \\ 
   &a_{\mu\nu}a^{\mu\nu},\qquad
  \left(a^\mu_\mu\right)^2,\qquad 
  a^\mu_\mu a^\nu a^\lambda a_{\nu\lambda}, \qquad
 a^\mu a_{\mu\nu}a^\lambda a^\nu_\lambda,\qquad 
 \left(a^\mu a^\nu a_{\mu\nu}\right)^2, \label{Q.bb}
\end{align}
multiplied by functions of $\phi$, $\mathcal{N}$, and $Z$.

At this stage, one may use the relation
\begin{align}
    \nabla_\mu\left[B(\phi,\mathcal{N},Z)a^\mu V\right]
    =Ba^\mu V_\mu+Ba_\mu^\mu V+\mathcal{N}B_{,\mathcal{N}}ZV
    +2B_{,Z}a^\mu a^\nu a_{\mu\nu}V
\end{align}
to remove one of the terms listed above.
We will remove this redundancy later.

Let us next consider the terms involving the curvature tensors.
In addition to
\begin{align}
    f(\phi,\mathcal{N}, Z)\mathcal{R},
\end{align}
one could add
\begin{align}
  n^\mu n^\nu\mathcal{R}_{\mu\nu}, \qquad
  a^\mu n^\nu\mathcal{R}_{\mu\nu},\qquad
  a^\mu a^\nu \mathcal{R}_{\mu\nu},\qquad 
  n^\mu a^\nu n^\rho a^\sigma \mathcal{R}_{\mu\nu\rho\sigma}, \label{R}
\end{align}
multiplied by functions of $\phi$, $\mathcal{N}$, and $Z$,
where $\mathcal{R}_{\mu\nu\rho\sigma}$ 
is the Riemann tensor and
$\mathcal{R}_{\mu\nu}$ is the Ricci tensor. 
However, all four terms in Eq.~\eqref{R} 
can be expressed as combinations of those appearing 
in Eq.~\eqref{Q.bb}, and hence they can be excluded 
from the building blocks.

Summarizing the above, our starting action in the covariant form is given by
\begin{align}
    S^{(\text{cov})}[g_{\mu\nu},\phi]=\int \D^4 x \sqrt{-g}
    \mathcal{L}^{(\text{cov})}
    , \label{eq:action_cov}
\end{align}
with
\begin{align}
        \mathcal{L}^{(\text{cov})}&=g_1 K+g_2a^\mu a^\nu K_{\mu\nu}
        +g_3V+g_4a^\mu V_\mu +g_5a^\mu a^\nu a_{\mu\nu}
        \notag \\ &\quad 
        +f\mathcal{R}
        +b_1 K_{\mu\nu} K^{\mu\nu}+b_2  K^2 
        +b_3  Ka^\mu a^\nu K_{\mu\nu}
        +b_4 a^\mu K_{\mu\nu}a^\lambda K_\lambda^\nu
        +b_5(a^\mu a^\nu K_{\mu\nu})^2
        \notag \\ &\quad 
        +c_1V^2+c_2V_\mu V^\mu+c_3(a^\mu V_\mu)^2
        +c_{4} V a^\mu V_\mu
        +c_5 KV+c_{6} a^\mu a^\nu K_{\mu\nu}V
        \notag \\ &\quad 
        +c_{7}a^\mu K_{\mu\nu}V^\nu
        +c_{8} Ka^\mu V_\mu
        +c_{9}a^\mu a^\nu a^\lambda K_{\mu\nu}V_\lambda
        \notag \\ & \quad 
        +d_0
        +d_1a_{\mu\nu} a^{\mu\nu}+d_2 (a_\mu^\mu)^2
        +d_3a_\mu^\mu a^\nu a^\lambda a_{\nu\lambda} 
        +d_4a^\mu a_{\mu\nu}a^\lambda a_{\lambda}^\nu+d_5 (a^\mu a^\nu a_{\mu\nu})^2
        \notag \\ &\quad 
        +h_{1}  K_{\mu\nu}a^{\mu\nu}
        +h_{2} Ka_\mu^\mu
        +h_{3}a^\mu a^\nu K_{\mu\nu} a_\lambda^\lambda
        +h_{4} Ka^\mu a^\nu a_{\mu\nu}
        +h_{5}a^\mu a^\nu K_{\mu\nu} a^\lambda a^\rho a_{\lambda\rho}
        +h_{6}a^\mu K_{\mu\nu}a^\lambda a_{\lambda}^\nu
        \notag \\ & \quad 
        +h_{7}V a_\mu^\mu
        +h_{8} Va^\mu a^\nu a_{\mu\nu}
        +h_{9}a_\mu^\mu a^\nu V_\nu 
        +
        h_{10}a^\mu a_{\mu\nu}V^\nu
        +h_{11}a^\mu a^\nu a^\lambda a_{\mu\nu}V_\lambda.
        \label{eq:covariant-Lagrangian}
\end{align}
In the following analysis, we drop the $\phi$-dependence of the coefficients
for simplicity
as it is not relevant to the degenerate structure of the theory.
Therefore, the coefficients are understood as functions of $\mathcal{N}$ (or $X$) and $Z$.
One could add the term $A_0(\mathcal{N},Z)$ to the above,
but we do not do so for the same reason.

\subsection{Unitary gauge action}

To apply the results of the Hamiltonian analysis~\cite{Gao:2018znj},
we perform the ADM decomposition of the covariant action introduced in the previous subsection,
which is straightforward because the action is already cast in a form suitable for that purpose.

Since we require that the scalar field $\phi$ acquires a timelike 
gradient, the time coordinate is allowed to be 
identified with the scalar field itself, i.e., $t=\phi$. 
This gauge choice is called the unitary gauge.
The unitary gauge action can be obtained simply by substituting
\begin{align}
        \mathcal{N}&\to N,
        \\
        \mathcal{P}_{\mu\nu}&\to \gamma_{ij},
        \\
        a_\mu &\to a_i:=\partial_i \ln N,
        \\ 
        V&\to V:=\frac{1}{N}\left(\partial_t N-N^i\partial_iN\right),
        \\ 
         K_{\mu\nu}&\to K_{ij}:=\frac{1}{2N}\left(\partial_t\gamma_{ij}-D_iN_j-D_jN_i\right),
        \\ 
        a_{\mu\nu}&\to a_{ij}:=D_ia_j,
        \\ 
        V_\mu&\to V_i:=\partial_iV,
\end{align}
and
\begin{align}
        \mathcal{R}- K_{\mu\nu} K^{\mu\nu}
        + K^2-2\nabla_\mu\left( Kn^\mu-a^\mu\right)
        &\rightarrow R^{(3)},\label{eq:transformation-curvature}
\end{align}
into the covariant action~\eqref{eq:action_cov}.
Here, $N$ is the lapse function, $N_i$ is the shift vector,
$\gamma_{ij}$ is the three-dimensional spatial metric on constant-time hypersurfaces,
$D_i$ is the covariant derivative compatible with $\gamma_{ij}$,
and $R^{(3)}$ is the three-dimensional Ricci scalar.
By using Eq.~\eqref{eq:transformation-curvature},
the curvature term $f\mathcal{R}$ in Eq.~\eqref{eq:covariant-Lagrangian}
can be rewritten, up to a total divergence, as
\begin{align}
        f \mathcal{R}
        &=fR^{(3)}
        +f\left( K_{\mu\nu} K^{\mu\nu}
        - K^2\right)
        \notag \\ & \quad 
        -4f_{,X}(-X)^{3/2} KV 
        -4f_{,Z} K\left(\sqrt{-X}a^\mu V_\mu-a^{\mu}a^\nu K_{\mu\nu}\right)
        -4f_{,X}XZ +4f_{,Z}a^\mu a^\nu a_{\mu\nu}.
\end{align}
This suggests that it is convenient to introduce
\begin{align}
        &\tilde b_1:=b_1+f,\qquad \tilde{b}_2:=b_2-f,
        \qquad 
        \tilde b_3:=b_3+4f_{,Z},
        \qquad 
        \tilde c_5:=c_5-4f_{,X}(-X)^{3/2},
        \notag \\  &
        \tilde c_8:=c_8-4f_{,Z}(-X)^{1/2},
        \qquad 
        \tilde d_0=d_0-4f_{,X}XZ,\qquad 
        \tilde g_{5}=g_{5}+4f_{,Z}.
\end{align}
In the unitary gauge, the action is thus given by
\begin{align}
    S^{(\text{u.g})}[N,N_i,\gamma_{ij}]=\int \D t \D^3x \sqrt{\gamma}N
    \mathcal{L}^{(\text{u.g.})}, \label{eq:action_UG}
\end{align}
with
\begin{align}
        \mathcal{L}^\text{(u.g.)}&=g_1K+g_2a^ia^jK_{ij}+g_3V+g_4a^iV_i 
        +\mathcal{L}_{\text{kin}}
        \notag \\ & \quad 
        +h_1K_{ij}a^{ij}+h_2Ka_i^i +h_3a^ia^jK_{ij}a_k^k +
        h_4Ka^ia^ja_{ij}
        \notag \\ & \quad 
        +h_5a^ia^jK_{ij}a^ka^la_{kl} 
        +h_6a^iK_{ij}a^ka_{k}^j 
        +h_7Va_i^i +h_8Va^ia^ja_{ij}
        \notag \\ & \quad 
        +h_9a^i_ia^jV_j
        +h_{10}a^ia_{ij}V^j 
        +h_{11}a^ia^ja_{ij}a^kV_k+\mathcal{L}_0,
        \label{eq:unitary-gauge-Lagrangian}
\end{align}
where $\mathcal{L}_{\text{kin}}$ is given by
\begin{align}
    \mathcal{L}_{\text{kin}}&=
    \tilde{b}_1K_{ij}K^{ij}+\tilde{b}_2K^2 
        +\tilde{b}_3 Ka^ia^jK_{ij}+b_4a^iK_{ij}a^kK^j_k 
        +b_5(a^ia^jK_{ij})^2 
        \notag \\ & \quad 
        +c_1V^2+c_2V_iV^i+c_3(a^iV_i)^2+c_4Va^iV_i
        +\tilde{c}_5KV+c_6a^ia^jK_{ij}V
        \notag \\ & \quad 
        +c_7a^iK_{ij}V^j 
        +\tilde{c}_8Ka^iV_i+c_9a^ia^jK_{ij}a^kV_k ,
\end{align}
and
$\mathcal{L}_0$ is the collection of terms that are independent of
$K_{ij}$, $V$, and $V_i$, and hence it is not relevant to the analysis in the following sections.

Let us introduce here the terminology used in this paper.
A quantity $T^{i_1i_2\dots}_{j_1j_2\dots}$ is said to be of \textit{type I} if
it is build solely from $N$, $a_i$, and $\gamma_{ij}$,
while it is said to be of \textit{type II} if it contains other tensors such as $K_{ij}$, $a_{ij}$, $V$, and $V_i$.

For later convenience, we introduce the following type I quantities
\begin{align}
    \mathcal{K}^{ij}&:=g_1\gamma^{ij}+g_2a^ia^j,\label{def:calKij}
    \\
    \mathcal{K}^{ij,kl}&:=
    \tilde{b}_1\left(\gamma^{ik}\gamma^{jl}+\gamma^{il}\gamma^{jk}\right)
    +2\tilde{b}_2\gamma^{ij}\gamma^{kl}
    +\tilde{b}_3(\gamma^{ij}a^k a^l+\gamma^{kl}a^i a^j)
    \notag \\ &\quad 
    +\frac{b_4}{2}\left(a^ia^k\gamma^{jl}+a^ia^l\gamma^{jk}
    +a^ja^k\gamma^{il}+a^ja^l\gamma^{ik}\right)
    +2b_5a^ia^ja^ka^l,
    \label{def:calKijkl}
    \\
    \mathcal{C}_2^{ij}&:=c_2\gamma^{ij}+c_3a^ia^j,
    \\
    \mathcal{C}_5^{ij}&:=\tilde c_5\gamma^{ij}+c_6a^ia^j,
    \\
    \mathcal{C}_7^{ij,k}&:=c_7a^{(i}\gamma^{j)k}+\tilde c_8\gamma^{ij}a^k+c_9a^ia^ja^k,
    \\
    \mathcal{H}_1^{ij,kl}&:=\frac{h_1}{2}\left(\gamma^{ik}\gamma^{jl}+
        \gamma^{il}\gamma^{jk}\right)+h_2\gamma^{ij}\gamma^{kl}+h_3 a^ia^j\gamma^{kl} 
        +h_4\gamma^{ij}a^ka^l +h_5a^ia^ja^ka^l 
        \notag \\ & \quad 
        +\frac{h_6}{4}\left(a^ia^k\gamma^{jl}+a^ia^l\gamma^{jk}
    +a^ja^k\gamma^{il}+a^ja^l\gamma^{ik}\right),
    \label{def:cal H1}\\ 
    \mathcal{H}_7^{ij}&:=h_7\gamma^{ij}+h_8a^ia^j,
    \\
    \mathcal{H}_9^{ij,k}&:=h_9\gamma^{ij}a^k+h_{10}a^{(i}\gamma^{j)k}+h_{11}a^ia^ja^k.\label{def:cal H9}
\end{align}
Using these quantities,
one can rewrite the unitary gauge Lagrangian into a compact form as
\begin{align}
    \mathcal{L}^{(\text{u.g.})}&=\mathcal{K}^{ij}K_{ij}+g_3V+g_4a^iV_i
    +\mathcal{L}_{\text{kin}}
    \notag \\ & \quad 
    +\mathcal{H}_1^{ij,kl}K_{ij}a_{kl}+\mathcal{H}_{7}^{ij}a_{ij}V
    +\mathcal{H}_{9}^{ij,k}a_{ij}V_k+\mathcal{L}_0,
    \label{eq:unitary-gauge-Lagrangian-reduced}
\end{align}
with
\begin{align}
    \mathcal{L}_{\text{kin}}&=
    \frac{1}{2} \mathcal{K}^{ij,kl}K_{ij}K_{kl}+c_1V^2+c_4Va^iV_i
    +\mathcal{C}_2^{ij}V_iV_j
    +\mathcal{C}_5^{ij}K_{ij}V+\mathcal{C}_7^{ij,k}K_{ij}V_k.
\end{align}

One may notice that the term $g_4 a^iV_i$ is redundant because it can be recast into some of the other terms by integration by parts.
Indeed, by adding the total divergence term
\begin{align}
    \int\D t\D^3x \sqrt{\gamma}N\cdot
    N^{-1}D_i\left[Ng_*(N,Z)a^iV\right] \label{eq:total-div}
\end{align}
to the above action, the coefficients $g_3$, $g_4$, and $\mathcal{H}_7^{ij}$ are shifted as follows:
\begin{align}
    g_3&\to g_3+\frac{\partial(Ng_*)}{\partial N}Z,
    \\ 
    g_4&\to g_4+g_*,
    \\
    \mathcal{H}_7^{ij}&\to \mathcal{H}_7^{ij}+g_*\gamma^{ij}+2\frac{\partial g_*}{\partial Z}a^ia^j.
\end{align}
This is nothing but the redundancy we have already mentioned in Sec.~\ref{subsec:bb}, and
we revisit this in Sec.~\ref{sec:fi=0}.

\section{The degeneracy conditions}\label{sec:degeneracy}

Our starting action introduced in the previous section contains the kinetic terms for both the spatial metric, $\gamma_{ij}$, and the lapse function, $N$, 
and also includes the terms mixing temporal and spatial derivatives.
The action is of the form to which the results of the rigorous Hamiltonian analysis performed in Ref.~\cite{Gao:2018znj} can be applied as it is,
showing that there are four propagating degrees of freedom in general,
i.e., there is one extra degree of freedom on top of one scalar and two tensor degrees of freedom.
It has been shown that two independent sets of conditions are required for removing this extra degree of freedom, namely the degeneracy and consistency conditions~\cite{Motohashi:2014opa, Motohashi:2016ftl, Motohashi:2017eya, Motohashi:2018pxg,Gao:2018znj}.\footnote{Imposing the degeneracy conditions, which require the kinetic matrix to be degenerate, leads to primary constraints.
In addition, when the action contains terms mixing temporal and spatial
derivatives, the consistency conditions, which 
ensure that the secondary constraints arise from 
the time preservation of the primary constraints, 
must be imposed. This allows us to eliminate the extra degree of freedom.
It should be noted that, for actions with higher-order derivatives, it was pointed out in Ref.~\cite{Motohashi:2014opa} that degeneracy alone is not sufficient, and the necessary additional conditions were derived for general point-particle systems in Refs.~\cite{Motohashi:2016ftl, Motohashi:2017eya, Motohashi:2018pxg}. Reference~\cite{Gao:2018znj} formulated these conditions in the framework of spatially covariant gravity.}
In this section, we derive the consequences of requiring the degeneracy conditions.

To begin with, we define the following object,
\begin{align}
    \mathcal{G}_{ij,kl}(x,y)=\frac{G_{ij,kl}(x)}{N\sqrt{\gamma}}\delta^{(3)}(x-y),
\end{align}
where $G_{ij,kl}$ is the inverse of $\mathcal{K}^{ij,kl}$ satisfying
\begin{align}
    \mathcal{K}^{ij,mn}G_{mn,kl}=\frac{1}{2}\left(\delta^i_k\delta^j_l
    +\delta^i_l\delta^j_k\right).\label{eq:KG-dddd}
\end{align}
It is shown in Appendix~\ref{app:inverse-of-calK} that such $G_{ij,kl}$ indeed exists.\footnote{In the following, we restrict attention to the branch in which the
metric-sector kinetic matrix $\mathcal K^{ij,kl}$ is invertible. The
explicit invertibility conditions are given in Eq.~\eqref{eq:assumption_inverse}.}
The degeneracy conditions are derived from~\cite{Gao:2018znj}
\begin{align}
    \mathcal{D}(x,y)&:=
    \frac{\delta^2S}{\delta V(x)\delta V(y)}-
    \int\D^3z\int\D^3w\frac{\delta^2S}{\delta V(x)\delta K_{ij}(z)}
    \mathcal{G}_{ij,kl}(z,w)\frac{\delta^2S}{\delta K_{kl}(w)\delta V(y)}
    =0.
\end{align}
Here, the variation is taken by treating $K_{ij}$ and $V$ as independent
variables.
It turns out that $\mathcal{D}(x,y)$ contains the delta function and its derivatives in general:
\begin{align}
    \mathcal{D}(x,y)=\sqrt{\gamma}\mathcal{D}_{(0)}(x)\delta^{(3)}(x-y)
    +\frac{\partial^2}{\partial x^i\partial y^j}\left[\sqrt{\gamma}\mathcal{D}_{(2)}^{ij}(x)\delta^{(3)}(x-y)\right],
\end{align}
where 
\begin{align}
    \mathcal{D}_{(0)}&=N\left(2c_1-
    \mathcal{C}_5^{ij}G_{ij,kl}\mathcal{C}_5^{kl}\right)
    -D_m\left[N\left(c_4a^m-\mathcal{C}_5^{ij}G_{ij,kl}\mathcal{C}_7^{kl,m}\right)\right],
    \\ 
    \mathcal{{D}}_{(2)}^{mn}&=N\left(
        2\mathcal{C}_2^{mn}-\mathcal{C}_7^{ij,m}G_{ij,kl}\mathcal{C}_7^{kl,n}
    \right).
\end{align}
The first and second variations of the action used to derive the above expressions are presented in Appendix~\ref{app:varDS}.

To deal with the derivatives of the delta function,
we introduce arbitrary smooth test functions $\varphi(x)$ 
and $\psi(x)$ with compact support.
Then, we consider the integral
\begin{align}
    0=\int\D^3x\int\D^3y \varphi(x)\psi(y)\mathcal{D}(x,y)
    =\int\D^3x\sqrt{\gamma} \varphi\psi\mathcal{D}_{(0)}(x)
    +\int\D^3x\sqrt{\gamma}\varphi_{|i}\psi_{|j}\mathcal{D}_{(2)}^{ij}(x).
\end{align}
Here, $\varphi_{|i}$ stands for $D_i\varphi$.
This leads to the conditions
\begin{align}
    \mathcal{D}_{(0)}=\mathcal{D}_{(2)}^{ij}=0.
\end{align}

A type I quantity $T^i$ must be of the form $T^i=T(N,Z)a^i$,
and therefore $T^i=(T^ja_j)a^i/Z$,
which allows us to write $\mathcal{D}_{(0)}$ in the form 
\begin{align}
    \mathcal{D}_{(0)}=2N\mathcal{C}_1-D_m\left(N\mathcal{C}_4a^m\right),
\end{align}
where
$\mathcal{C}_1:=c_1-\mathcal{C}_5^{ij}G_{ij,kl}\mathcal{C}_5^{kl}/2$
and 
$\mathcal{C}_4:=c_4-\mathcal{C}_5^{ij}G_{ij,kl}\mathcal{C}_7^{kl,m}a_m/Z$.
By requiring that $\mathcal{D}_{(0)}=0$ for any configuration of $a_i$,
we obtain $\mathcal{C}_1=\mathcal{C}_4=0$.
The degeneracy conditions thus yield
\begin{align}
    c_1&=\frac{1}{2}\mathcal{C}_5^{ij}G_{ij,kl}\mathcal{C}_5^{kl},
    \\ 
    c_4&=\frac{1}{Z}\mathcal{C}_5^{ij}G_{ij,kl}\mathcal{C}_7^{kl,m}a_m,
    \\
    \mathcal{C}_2^{ij}&=\frac{1}{2}\mathcal{C}_7^{mn,i}G_{mn,kl}\mathcal{C}_7^{kl,j}.
    \label{eq:c2mn=}
\end{align}
Using these relations, $\mathcal{L}_{\text{kin}}$ can be cast into a completed-square form as
\begin{align}
    \mathcal{L}_{\text{kin}}
    =\frac{1}{2} \mathcal{K}^{ij,kl}Q_{ij}Q_{kl},
\end{align}
where 
\begin{align}
    Q_{ij}:=K_{ij}+\mathcal{U}_{ij}V+\frac{1}{2}\mathcal{V}_{ij}^kV_k,
\end{align}
with 
\begin{align}
    \mathcal{U}_{ij}&:=G_{ij,mn}\mathcal{C}_5^{mn},
    \\ 
    \mathcal{V}_{ij}^k&:=2G_{ij,mn}\mathcal{C}_7^{mn,k}.
\end{align}
At this stage, there are no further constraints on the forms of
$\mathcal{U}_{ij}$ and $\mathcal{V}_{ij}^k$.

Note in passing that Eq.~\eqref{eq:c2mn=} allows us to write $c_2$ and $c_3$ explicitly as
\begin{align}
    c_2&=\frac{1}{4}\mathcal{C}_7^{ij,m}G_{ij,kl}\mathcal{C}_7^{kl,n}
    \left(\gamma_{mn}-\frac{a_ma_n}{Z}\right),
    \\
    c_3&=-\frac{1}{4Z}\mathcal{C}_7^{ij,m}G_{ij,kl}\mathcal{C}_7^{kl,n}
    \left(\gamma_{mn}-\frac{3a_ma_n}{Z}\right).
\end{align}

\section{The consistency conditions}\label{sec:consistency}

We now move to derive the consequences obtained by imposing the consistency conditions.
In the course of the following derivation, we make use of the degeneracy conditions obtained in the previous section when appropriate.
Since the derivation is quite involved, a short summary of the consequences of the consistency conditions is provided in Sec.~\ref{subsec:summary} for those who are not interested in technical details.

The consistency conditions are derived from~\cite{Gao:2018znj}
\begin{align}
    \mathcal{F}(x,y)=0,
\end{align}
where
\begin{align}
   \mathcal{F}(x,y)&:=
   \frac{1}{N(y)}\frac{\delta^2 S}{\delta N(x)\delta V(y)}
   -\frac{1}{N(x)}\frac{\delta^2 S}{\delta N(y)\delta V(x)}\\ & \quad
   +\int \D^3z \, \Biggl\{\Biggl[-\frac{1}{2}
   \delta^{(3)}(x-z)\frac{1}{N^2(z)}
   \frac{\delta S}{\delta K_{ij}(z)}\nonumber \\ & \quad
   +\frac{1}{2}\frac{1}{N(z)}\frac{\delta^2 S}
   {\delta N(x)\delta K_{ij}(z)}
   -\frac{1}{N(x)}\frac{\delta^2 S}
   {\delta V(x) \delta \gamma_{ij}(z)}
   \Biggr]
   \mathcal{V}_{ij}(z,y)
   -(x \leftrightarrow y)\Biggr\} \nonumber\\ & \quad
   +\int \D^3x'  \int \D^3y' \,
   \mathcal{V}_{ij}(x',x)
   \left[\frac{1}{2N(y')}\frac{\delta^2 S}
    {\delta \gamma_{ij}(x')\delta K_{kl}(y')}
    -\frac{1}{2N(x')}\frac{\delta^2 S}
    {\delta \gamma_{kl}(y')\delta K_{ij}(x')}
    \right]\mathcal{V}_{kl}(y',y),
\end{align}
with
\begin{align}
    \mathcal{V}_{ij}(x,y)=
    \left(\mathcal{V}_{ij}^ka_k-2\mathcal{U}_{ij}\right)
    \delta^{(3)}(x-y)
    +\mathcal{V}_{ij}^{k}(x)\frac{\partial\delta^{(3)}(x-y)}{\partial y^k}.
\end{align}

As in the previous section,
we introduce arbitrary smooth test functions $\phi$ and $\psi$ with compact support
to deal with the derivatives of the delta function.
One can then write
\begin{align}
    \int \D^3x\int \D^3y\, \varphi(x)\psi(y)\mathcal{F}(x,y)& 
    =\int\D^3x\sqrt{\gamma}
    \left(\varphi_{|i}\psi-\varphi\psi_{|i}\right)\mathcal{F}_{(1)}^{i}(x)
    \notag \\ & \quad 
    +\int\D^3x\sqrt{\gamma}
    \left(\varphi_{|i}\psi_{|j}-\varphi_{|j}\psi_{|i}\right)\mathcal{F}_{(2)}^{ij}(x)
    \notag \\ & \quad 
    +\int\D^3x\sqrt{\gamma}
    \left(\varphi_{|ij}\psi_{|k}-\varphi_{|k}\psi_{|ij}\right)\mathcal{F}_{(3)}^{ij,k}(x),
    \label{eq:decompose-F}
\end{align}
where $\mathcal{F}_{(2)}^{ij}$ and $\mathcal{F}_{(3)}^{ij,k}$ are, respectively, antisymmetric and symmetric under the exchange of indices $i$ and $j$.
Using the degeneracy conditions, the explicit expressions are found to be
\begin{align}
    \mathcal{F}_{(1)}^i&=\mathcal{A}_{kl}^i\left(\mathcal{K}^{kl}+\mathcal{K}^{kl,mn}Q_{mn}
    +\mathcal{H}_1^{kl,mn}a_{mn}\right)+\mathcal{B}^{ij}_{kl}\mathcal{H}_1^{kl,mn}a_ja_{mn}
    +\mathcal{X}^{kl,i}a_{kl}+\mathcal{Y}^i,
    \\
    \mathcal{F}_{(2)}^{ij}&=-\frac{1}{2}\mathcal{B}^{ij}_{kl}\left(\mathcal{K}^{kl,mn}Q_{mn}
        +\mathcal{H}_1^{kl,mn}a_{mn}
        \right)+\mathcal{Z}^{ij,kl}a_{kl},
\end{align}
and 
\begin{align}
    \mathcal{F}_{(3)}^{ij,k}&=
    \frac{1}{N}\mathcal{E}^{ij,k}+\frac{1}{2}\mathcal{V}_{mn}^{(i}
    \left(
        2\mathcal{E}^{j)m,k}a^n-a^{j)}\mathcal{E}^{mn,k}
    \right),
\end{align}
where
\begin{align}
    \mathcal{A}_{kl}^i&:=
    \mathcal{U}_{mn}\frac{\partial\mathcal{V}_{kl}^i}{\partial\gamma_{mn}}
        -\mathcal{V}_{mn}^i\frac{\partial\mathcal{U}_{kl}}{\partial\gamma_{mn}}
        +\frac{1}{N}\left[
                \frac{\partial\mathcal{U}_{kl}}{\partial a_i}-\frac{1}{2}\frac{\partial}{\partial N}
                (N\mathcal{V}_{kl}^i)
        \right],
        \label{def:calA}
    \\
    \mathcal{B}^{ij}_{kl}&:=
    \frac{1}{N}\frac{\partial\mathcal{V}_{kl}^{[i}}{\partial a_{j]}}
                -\frac{\partial\mathcal{V}_{kl}^{[i}}{\partial\gamma_{mn}}\mathcal{V}_{mn}^{j]},
    \label{def:calB}
    \\ 
    \mathcal{X}^{kl,i}&:=\frac{1}{N}\left(
        \frac{\partial\mathcal{E}^{kl}}{\partial a_  i}-\frac{\partial\mathcal{E}^{i(k}}{\partial a_{l)}}
    \right)-\mathcal{V}_{mn}^i
    \left(
        \frac{\partial\mathcal{E}^{kl}}{\partial\gamma_{mn}}
        +\frac{1}{2}\gamma^{mn}\mathcal{E}^{kl}
        +
        a^n\frac{\partial \mathcal{E}^{m(k}}{\partial a_{l)}}+
        \mathcal{E}^{m(k}\gamma^{l)n}
        -\frac{1}{2}a^{(k}\frac{\partial\mathcal{E}^{mn}}{\partial a_{l)}}
        -\frac{1}{2}\mathcal{E}^{mn}\gamma^{kl}
    \right)
    \notag \\ & \quad
    -\frac{\partial\mathcal{E}^{kl,i}}{\partial N}+\frac{1}{N}\mathcal{E}^{i(k,l)}
    +\frac{a_j}{N}\left(
        \frac{\partial\mathcal{E}^{kl,i}}{\partial a_j}
        -\frac{\partial \mathcal{E}^{kl,j}}{\partial a_i}
        +\frac{\partial\mathcal{E}^{l)i,j}}{\partial a_{(k}}
    \right)
        -\frac{\partial\mathcal{U}_{mn}}{\partial a_{(k}}
        \left(
            2\mathcal{E}^{l)m,i}a^n-a^{l)}\mathcal{E}^{mn,i}
        \right)
    \notag \\ & \quad 
    +a_j\mathcal{V}_{mn}^i\left(
        \frac{\partial\mathcal{E}^{kl,j}}{\partial\gamma_{mn}}+\frac{1}{2}\gamma^{mn}
        \mathcal{E}^{kl,j}
    \right)
    +\frac{1}{2}\left(
        \mathcal{V}_{mn}^{(k}+a_j\frac{\partial\mathcal{V}_{mn}^j}{\partial a_{(k}}
    \right)\left(
        2\mathcal{E}^{l)m,i}a^n-a^{l)}\mathcal{E}^{mn,i}
    \right)
    \notag \\ & \quad 
    +\frac{\mathcal{V}_{mn}^i}{2}
    \left(
        2a^m\mathcal{E}^{n(k,l)}-\mathcal{E}^{mn,(k}a^{l)}+
        2a_j\gamma^{m(k}\mathcal{E}^{l)n,j}-a_j\gamma^{kl}\mathcal{E}^{mn,j}
        +2a_ja^m\frac{\partial\mathcal{E}^{l)n,j}}{\partial a_{(k}}
        -a_ja^{(k}\frac{\partial\mathcal{E}^{mn,j}}{\partial a_{l)}}
    \right),
    \\
    \mathcal{Y}^i&:=\frac{1}{N}\frac{\partial\mathcal{G}_3}{\partial a_i}
    -\mathcal{V}_{mn}^i\left(
        \frac{\partial\mathcal{G}_3}{\partial\gamma_{mn}}+\frac{1}{2}\gamma^{mn}\mathcal{G}_3
    \right)
    -\frac{a^i}{N}\frac{\partial(N\mathcal{G}_4)}{\partial N}
    +2\mathcal{U}_{mn}
    \left(
        a^i\frac{\partial\mathcal{G}_4}{\partial\gamma_{mn}}
        -\gamma^{i(m}a^{n)}\mathcal{G}_4+
        \frac{1}{2}a^i\gamma^{mn}\mathcal{G}_4
    \right)
    \notag \\ & \quad 
    -\frac{a_j}{N}\frac{\partial(N\mathcal{E}^{ij})}{\partial N}
    +\mathcal{U}_{mn}\left(
        2\mathcal{E}^{im}a^n-a^i\mathcal{E}^{mn}
    \right)
    -\frac{1}{2}\mathcal{V}^i_{mn}a_k
    \left[
        2\frac{\partial(N\mathcal{E}^{km})}{\partial N}a^n
        -\frac{\partial(N\mathcal{E}^{mn})}{\partial N}a^k
    \right]
    \notag \\ & \quad 
    +a_ja_k\left[\frac{\partial (N\mathcal{F}^{ij,k}_{(3)})}{\partial N}
    +\mathcal{F}^{ij,k}_{(3)}-\mathcal{F}^{kj,i}_{(3)}
    \right]
    -\frac{\partial(N\mathcal{U}_{mn})}{\partial N}a_k
    \left(2a^m\mathcal{E}^{nk,i}-a^k\mathcal{E}^{mn,i}\right)
    ,
    \\ 
    \mathcal{Z}^{ij,kl}&:=
            -\frac{1}{N}\frac{\partial\mathcal{E}^{kl,[i}}{\partial a_{j]}}
            +\left(
            \frac{\partial\mathcal{E}^{kl,[i}}{\partial\gamma_{mn}}
            +\frac{1}{2}\gamma^{mn}\mathcal{E}^{kl,[i}
            \right)
            \mathcal{V}_{mn}^{j]}
            -
            \frac{1}{2}\left(
                2a^{m}\mathcal{E}^{n(k,[i}-a^{(k}\mathcal{E}^{mn,[i}
        \right)\frac{\partial\mathcal{V}_{mn}^{j]}}{\partial a_{l)}},
\end{align}
with
\begin{align}
    \mathcal{G}_3&:=g_3-\mathcal{U}_{ij}\mathcal{K}^{ij},
    \\ 
    \mathcal{G}_4&:=g_4 - \frac{1}{2Z}a_i\mathcal{V}_{mn}^i\mathcal{K}^{mn},
    \\
    \mathcal{E}^{ij}&:=\mathcal{H}_7^{ij}-\mathcal{U}_{mn}\mathcal{H}_1^{mn,ij},
    \\
    \mathcal{E}^{ij,k}&:=\mathcal{H}_9^{ij,k}-\frac{1}{2}\mathcal{V}_{mn}^k\mathcal{H}_1^{mn,ij}.
    \label{def:calEijk}
\end{align}
To derive the above expressions, we used the first and second variations of the action presented in Appendix~\ref{app:varDS}.
The quantities~\eqref{def:calA}--\eqref{def:calEijk} are of type I.

In Eq.~\eqref{eq:decompose-F}, the test functions 
$\varphi$ and $\psi$ appear in the following three 
combinations:
\begin{align*}
    \varphi_{|i}\psi-\varphi\psi_{|i}, \qquad 
    \varphi_{|i}\psi_{|j}-\varphi_{|j}\psi_{|i}, \qquad 
    \varphi_{|ij}\psi_{|k}-\varphi_{|k}\psi_{|ij}.     
\end{align*}
These three patterns can be transformed 
into one another by integration by parts.
Therefore, there remains some redundancy in this decomposition, and one cannot simply conclude that $\mathcal{F}_{(1)}^i=0$, $\mathcal{F}_{(2)}^{ij}=0$, and $\mathcal{F}_{(3)}^{ij,k}=0$ follow from $\mathcal{F}=0$.
We can remove the redundancy and extract the appropriate set of equations from $\mathcal{F}=0$ as follows.

Let us split $\mathcal{F}_{(3)}^{ij,k}$ into the totally symmetric component and the rest as
\begin{align}
    \mathcal{F}_{(3)}^{ij,k}=\mathcal{F}_{(3)}^{(ij,k)}+\widetilde{\mathcal{F}}_{(3)}^{ij,k},
\end{align}
where
\begin{align}
    \mathcal{F}_{(3)}^{(ij,k)}&:=\frac{1}{3}\left(
        \mathcal{F}_{(3)}^{ij,k}+\mathcal{F}_{(3)}^{jk,i}+\mathcal{F}_{(3)}^{ki,j}
    \right),
    \\
    \widetilde{\mathcal{F}}_{(3)}^{ij,k}&:=\frac{1}{3}\left(
        2\mathcal{F}_{(3)}^{ij,k}-\mathcal{F}_{(3)}^{jk,i}-\mathcal{F}_{(3)}^{ki,j}
    \right).
\end{align}
We have the relation
\begin{align}
    \widetilde{\mathcal{F}}^{ij,k}_{(3)}+\widetilde{\mathcal{F}}^{jk,i}_{(3)}
    +\widetilde{\mathcal{F}}^{ki,j}_{(3)}=0,
\end{align}
using which we obtain
\begin{align}
    \int\D^3x\sqrt{\gamma}
    \left(\varphi_{|ij}\psi_{|k}-\varphi_{|k}\psi_{|ij}\right)\widetilde{\mathcal{F}}^{ij,k}_{(3)}
    &=\frac{1}{3}\int\D^3x\sqrt{\gamma}
    \left(\varphi_{|jk}\psi_{|i}-\varphi_{|i}\psi_{|jk}\right)\widetilde{\mathcal{F}}^{jk,i}_{(3)}
    \notag \\ & \quad 
    +\frac{1}{3}\int\D^3x\sqrt{\gamma}
    \left(\varphi_{|ki}\psi_{|j}-\varphi_{|j}\psi_{|ki}\right)\widetilde{\mathcal{F}}^{ki,j}_{(3)}
    \notag \\ & \quad 
    +\frac{1}{3}\int\D^3x\sqrt{\gamma}
    \left(\varphi_{|ij}\psi_{|k}-\varphi_{|k}\psi_{|ij}\right)\widetilde{\mathcal{F}}^{ij,k}_{(3)}
    \notag \\ & =
    \frac{2}{3}\int\D^3x\sqrt{\gamma}
    \left(\varphi_{|i}\psi_{|j}-\varphi_{|j}\psi_{|i}\right)_{|k}\widetilde{\mathcal{F}}^{ki,j}_{(3)}
    \notag \\ & = -
    \frac{2}{3}\int\D^3x\sqrt{\gamma}
    \left(\varphi_{|i}\psi_{|j}-\varphi_{|j}\psi_{|i}\right)D_k\widetilde{\mathcal{F}}^{ki,j}_{(3)}
    \notag \\ & =-\frac{1}{3}\int\D^3x\sqrt{\gamma}
    \left(\varphi_{|i}\psi_{|j}-\varphi_{|j}\psi_{|i}\right)D_k\left(
    \mathcal{F}_{(3)}^{ki,j}-\mathcal{F}_{(3)}^{kj,i}
    \right).
\end{align}
Thus, this contribution can be incorporated into the second line of Eq.~\eqref{eq:decompose-F},
giving
\begin{align}
    \int\D^3x\sqrt{\gamma}
    \left(\varphi_{|i}\psi_{|j}-\varphi_{|j}\psi_{|i}\right)
    \left[\mathcal{F}_{(2)}^{ij}-\frac{1}{3}D_k \left(\mathcal{F}_{(3)}^{ki,j}
    -\mathcal{F}_{(3)}^{kj,i}\right)
    \right].
\end{align}

Upon integration by parts, we further obtain
\begin{align}
    &\int\D^3x\sqrt{\gamma}
    \left(\varphi_{|i}\psi_{|j}-\varphi_{|j}\psi_{|i}\right)
    \left[\mathcal{F}_{(2)}^{ij}-\frac{1}{3}D_k \left(\mathcal{F}_{(3)}^{ki,j}
    -\mathcal{F}_{(3)}^{kj,i}\right)
    \right]
    \notag \\ & 
    =-
    \int\D^3x\sqrt{\gamma}
    \left(\varphi_{|i}\psi-\varphi\psi_{|i}\right)D_j
    \left[\mathcal{F}_{(2)}^{ij}-\frac{1}{3}D_k \left(\mathcal{F}_{(3)}^{ki,j}
    -\mathcal{F}_{(3)}^{kj,i}\right)
    \right].
\end{align}
Equation~\eqref{eq:decompose-F} can thus be rewritten as
\begin{align}
    \int \D^3x\int \D^3y\, \varphi(x)\psi(y)\mathcal{F}(x,y)& 
    =\int\D^3x\sqrt{\gamma}
    \left(\varphi_{|i}\psi-\varphi\psi_{|i}\right)
    \left\{\mathcal{F}_{(1)}^{i}
    -D_j
    \left[\mathcal{F}_{(2)}^{ij}-\frac{1}{3}D_k \left(\mathcal{F}_{(3)}^{ki,j}
    -\mathcal{F}_{(3)}^{kj,i}\right)
    \right]
    \right\}
    \notag \\ & \quad 
    +\int\D^3x\sqrt{\gamma}
    \left(\varphi_{|ij}\psi_{|k}-\varphi_{|k}\psi_{|ij}\right)\mathcal{F}_{(3)}^{(ij,k)}.\label{eq:reduced-decompose-F}
\end{align}
No further reduction via integration by parts is possible for the symmetric part $\mathcal{F}_{(3)}^{(ij,k)}$.
The consistency conditions thus end up with the following two equations:
\begin{align}
    \mathcal{F}^i:=
    \mathcal{F}_{(1)}^{i}
    -D_j
    \left[\mathcal{F}_{(2)}^{ij}-\frac{1}{3}D_k \left(\mathcal{F}_{(3)}^{ki,j}
    -\mathcal{F}_{(3)}^{kj,i}\right)
    \right]=0,
    \label{eq:cond-F1}
\end{align}
and 
\begin{align}
    \mathcal{F}_{(3)}^{(ij,k)}=0.
    \label{eq:cond-F3}
\end{align}
Let us investigate the implications of Eqs.~\eqref{eq:cond-F1} and~\eqref{eq:cond-F3}.

\subsection{The consistency condition $\mathcal{F}_{(3)}^{(ij,k)}=0$}

Since $\mathcal{E}^{ij,k}$ and $\mathcal{V}_{ij}^k$ are type I quantities, they can be written in the form
\begin{align}
    \mathcal{E}^{ij,k}&=E_1(N,Z)\gamma^{ij}a^k+E_2(N,Z)\gamma^{k(i}a^{j)}+
    E_3(N,Z)a^ia^ja^k,\label{ansatz:cal Eijk}
    \\ 
    \mathcal{V}_{ij}^k&=
    v_1(N,Z)\gamma_{ij}a^k+v_2(N,Z)a_{(i}\delta_{j)}^k
    +v_3(N,Z)a_ia_ja^k.\label{ansatz:cal Vijk}
\end{align}
Substituting these expressions into $\mathcal{F}_{(3)}^{ij,k}$, we obtain
\begin{align}
    \mathcal{F}_{(3)}^{(ij,k)}=
    \mathcal{F}_{(3)}^{\gamma a}\gamma^{(ij}a^{k)}+
    \mathcal{F}_{(3)}^{aaa}a^ia^ja^k,
\end{align}
where
\begin{align}
    \mathcal{F}_{(3)}^{\gamma a}&:=\frac{E_1+E_2}{N}+\frac{1}{2}Z\left[
        v_2E_1
    +\left(v_1+v_2+Zv_3\right)E_2
    \right], \label{def:A}
    \\
    \mathcal{F}_{(3)}^{aaa}&:=\frac{E_3}{N}+\frac{1}{2}\left[
        (-v_1+Zv_3)E_1
        +Z(v_1+v_2+Zv_3)E_3
    \right].\label{def:B}
\end{align}
Therefore, the consistency condition~\eqref{eq:cond-F3} leads to 
$\mathcal{F}_{(3)}^{\gamma a}=\mathcal{F}_{(3)}^{aaa}=0$,
which can be used to express $E_2$ and $E_3$ in terms of $E_1$, $v_1$, $v_2$, and $v_3$.
We thus arrive at
\begin{align}
    \mathcal{E}^{ij,k}=E_1(N,Z)\left[
    \gamma^{ij}a^k-\frac{1+NZv_2/2}{1+NZ(v_1+v_2+Zv_3)/2}a^{(i}\gamma^{j)k} 
    +\frac{N(v_1-Zv_3)/2}{1+NZ(v_1+v_2+Zv_3)/2}a^ia^ja^k\right]. \label{sol:Eijk}
\end{align}
Using this result, we obtain
\begin{align}
    \mathcal{F}^{ij,k}_{(3)}=\widetilde{\mathcal{F}}^{ij,k}_{(3)}
    =\frac{E_1}{N}\left(1+\frac{NZv_2}{2}\right)
    \left(\gamma^{ij}a^k-a^{(i}\gamma^{j)k}\right).
\end{align}

\subsection{The consistency condition $\mathcal{F}^i=0$}\label{sec:fi=0}

Let us begin with splitting the type II quantity $\mathcal{F}^i$ into two pieces as
\begin{align}
    \mathcal{F}^i=\mathcal{F}^i_{Q}+\mathcal{F}^i_{\text{others}},
    \label{eq:F_Q+F_others}
\end{align}
where $\mathcal{F}^i_Q$ depends on
$Q_{ij}$, while $\mathcal{F}^i_{\text{others}}$ includes neither
$K_{ij}$ nor $V$.
In order for $\mathcal{F}^i=0$ to hold for any configuration of $K_{ij}$ and $V$, it is required that each piece vanishes: $\mathcal{F}_Q^i=\mathcal{F}^i_{\text{others}}=0$.
Explicitly, we have
\begin{align}
    \mathcal{F}_Q^i=\mathcal{A}^i_{kl}\mathcal{K}^{kl,mn}Q_{mn}
    +\frac{1}{2}D_j\left(\mathcal{B}^{ij}_{kl}\mathcal{K}^{kl,mn}Q_{mn}\right),
\end{align}
and thus
\begin{align}
    \mathcal{A}_{kl}^i&=0,\label{eq:FiQ2}
    \\ 
    \mathcal{B}_{kl}^{ij}&=0,\label{eq:FiQ1}
\end{align}
must be satisfied.
These conditions put some restrictions on the possible form of $\mathcal{U}_{ij}$ and $\mathcal{V}_{ij}^k$.

To solve Eqs.~\eqref{eq:FiQ2} and~\eqref{eq:FiQ1}, we use, in addition to Eq.~\eqref{ansatz:cal Vijk}, the expression
\begin{align}
    \mathcal{U}_{ij}&=u_1(N,Z)\gamma_{ij}+u_2(N,Z)a_ia_j,
    \label{eq:ansatz:Uij}
\end{align}
and derive the differential equations for the five functions 
$u_1, u_2, v_1, v_2$, and $v_3$. Substituting Eq.~\eqref{ansatz:cal Vijk} to the definition~\eqref{def:calB} of $\mathcal{B}^{ij}_{mn}$, we obtain 
\begin{align}
    \mathcal{B}^{ij}_{mn}
    \propto
    \left\{
        v_2'-v_3+\frac{N}{2}
        \left[
            v_1v_2+Zv_2'\left(v_1+v_2+Zv_3\right)
        \right]
    \right\}a^{[i}\delta^{j]}_{(m}a_{n)}=0,
\end{align}
leading to
\begin{align}
    v_2'-v_3+\frac{N}{2}
        \left[
            v_1v_2+Zv_2'\left(v_1+v_2+Zv_3\right)
        \right]=0,
        \label{eq:uveq0}
\end{align}
where a prime denotes differentiation with respect to $Z$.
Similarly, it follows from Eq.~\eqref{def:calA} that the equation $\mathcal{A}_{ij}^k=0$ can be written in the form of
$(\dots)\gamma_{ij}a^k+(\dots)a_{(i}\delta_{j)}^k+(\dots)a_ia_ja^k$=0.
This yields the following three equations:
\begin{align}
        2u_1'-Nu_1v_1
        +NZ\left(
                -v_1'u_1+u_1'v_1-u_2v_1+u_1'v_2
        \right)
        +NZ^2\left(-v_1'u_2+u_1'v_3\right)
        &=\frac{1}{2}\frac{\partial(Nv_1)}{\partial N},
        \label{eq:uveq1}
        \\ 
        2u_2-Nu_1v_2-N Z v_2'(u_1+Zu_2)
        &=\frac{1}{2}\frac{\partial (Nv_2)}{\partial N},
        \label{eq:uveq2}
        \\ 
        2u_2'+N(u_2v_1-2u_1v_3)+NZ\left(
                -v_3' u_1+u_2'v_1+u_2'v_2-u_2v_3
        \right)
        +NZ^2(-v_3'u_2+u_2'v_3)
        &=\frac{1}{2}\frac{\partial (Nv_3)}{\partial N}.
        \label{eq:uveq3}
\end{align}
It can be shown that $u_1,\dots,v_3$ satisfying Eqs.~\eqref{eq:uveq0}--\eqref{eq:uveq3} can be expressed as 
\begin{align}
    2Nu_1&=\frac{\partial W_1/\partial\ln N}{1-\partial W_1/\partial\ln Z}
    \left(
        1-\frac{\partial W_2}{\partial\ln Z}
    \right)-\frac{\partial W_2}{\partial \ln N},\label{sol:u1}
    \\
    2NZu_2&=\frac{\partial W_1/\partial\ln N}{1-\partial W_1/\partial\ln Z}
    \frac{\partial W_2}{\partial\ln Z}
    +\frac{\partial W_2}{\partial \ln N},
    \\ 
    \frac{NZ}{2}v_1&=\frac{\partial W_1/\partial \ln Z-\partial W_2/\partial\ln Z}{1-\partial W_1/\partial \ln Z},
    \\ 
    \frac{NZ}{2}v_2&=e^{W_2}-1,
    \\ 
    \frac{NZ^2}{2}v_3&=1-e^{W_2}+
    \frac{\partial W_2/\partial\ln Z}{1-\partial W_1/\partial \ln Z}, \label{sol:u_1-v_3}
\end{align}
where $W_1$ and $W_2$ are arbitrary functions of $N$ and $Z$.\footnote{A sufficient set of conditions for a regular homogeneous limit is
that $W_1$ be smooth at $Z=0$, that $W_2$ be smooth and satisfy
$W_2(t,N,0)=0$, or equivalently,
$W_2=Z\mathcal W_2$ with $\mathcal W_2$ smooth at $Z=0$, and that
$1-Z\partial_ZW_1\neq0$ on the branch under consideration. Under these assumptions,
all the coefficients $u_1,u_2,v_1,v_2,v_3$ admit finite limits as $Z\to0$.}
The detailed derivation is given in Appendix~\ref{app:der1}.

Let us move to investigating the structure of the condition $\mathcal{F}_{\text{others}}^i=0$.
Setting $\mathcal{A}_{kl}^i=\mathcal{B}_{kl}^i=0$, we have 
\begin{align}
    \mathcal{F}_{\text{others}}^i&= \mathcal{X}^{kl,i}a_{kl}+\mathcal{Y}^i
    -D_j\left[
        \mathcal{Z}^{ij,kl}a_{kl}-\frac{1}{3}D_k
        \left(\mathcal{F}_{(3)}^{ki,j}-\mathcal{F}_{(3)}^{kj,i}\right)
    \right].
\end{align}
However, a direct calculation with the help of Eq.~\eqref{eq:uveq0} shows that
\begin{align}
    \mathcal{Z}^{ij,kl}a_{kl}-\frac{1}{3}D_k
    \left(\mathcal{F}_{(3)}^{ki,j}-\mathcal{F}_{(3)}^{kj,i}\right)
    =0.
\end{align}
Therefore, $\mathcal{F}_{\text{others}}^i=0$ indicates that
\begin{align}
    \mathcal{X}^{kl,i}&=0,\label{eq:X=0} \\
    \mathcal{Y}^i&=0. \label{eq:Y=0}
\end{align}
Since $\mathcal{X}^{ij,k}$ and $\mathcal{Y}^i$ are type I quantities, one can write
\begin{align}
    \mathcal{X}^{ij,k}&=\mathcal{X}_1(N,Z)\gamma^{ij}a^k
    +\mathcal{X}_2(N,Z)\gamma^{k(i}a^{j)}+
    \mathcal{X}_3(N,Z)a^ia^ja^k,
    \\ 
    \mathcal{Y}^i&=\mathcal{Y}(N,Z)a^i.
\end{align}
Similarly, $\mathcal{E}^{ij}$ can be written as
\begin{align}
    \mathcal{E}^{ij}=e_1(N,Z)\gamma^{ij}+e_2(N,Z)a^ia^j.
\end{align}
The consistency conditions derived from $\mathcal{F}_{\text{others}}^i=0$ are summarized as
\begin{align}
    \mathcal{X}_1=\mathcal{X}_2=\mathcal{X}_3=\mathcal{Y}=0,
\end{align}
where the explicit expressions for $\mathcal{X}_1$, $\mathcal{X}_2$, and $\mathcal{X}_3$ are given by
\begin{align}
    \mathcal{X}_1&=\frac{1}{N}\left[1+\frac{NZ}{2}\left(v_1+v_2+Zv_3\right)\right]
    \left(2e_1'-e_2-E_1\right)
    \notag \\ & \quad 
    -\frac{\partial E_1}{\partial N}-Z^2\left(v_1+v_2+Zv_3\right)E_1'
    -2Zu_2E_1,
    \label{eq:calx1}
    \\ 
    \mathcal{X}_2&=-\frac{1}{N}\left(1+\frac{NZv_2}{2}\right)
    \left(2e_1'-e_2-E_1\right)
    \notag \\ & \quad 
    +\frac{\partial}{\partial N}\left[
        \frac{(1+NZv_2/2)E_1}{1+NZ(v_1+v_2+Zv_3)/2}
    \right]+Z^2\left(v_1+v_2+Zv_3\right)\left[
        \frac{(1+NZv_2/2)E_1}{1+NZ(v_1+v_2+Zv_3)/2}
    \right]'
    \notag \\ & \quad 
    +2Z\left(u_1+Zu_2\right)'\left[
        \frac{1+NZv_2/2}{1+NZ(v_1+v_2+Zv_3)/2}
        \right]
    E_1,
    \\ 
    \mathcal{X}_3&=\frac{1}{2}\left(v_1-Zv_3\right)\left(2e_1'-e_2-E_1\right)
    \notag \\ & \quad 
    -\frac{\partial}{\partial N}\left[
        \frac{N(v_1-Zv_3)E_1/2}{1+NZ(v_1+v_2+Zv_3)/2}
    \right]
    -Z\left(v_1+v_2+Zv_3\right)\left[
        \frac{NZ(v_1-Zv_3)E_1/2}{1+NZ(v_1+v_2+Zv_3)/2}
    \right]'
    \notag \\ & \quad 
    -2(u_1+Zu_2)'\left[
        \frac{NZ(v_1-Zv_3)/2}{1+NZ(v_1+v_2+Zv_3)/2}
    \right]E_1+\left[
        2\left(u_1'-Zu_2'\right)-\frac{NZv_1(v_2+Zv_3)}{1+NZ(v_1+v_2+Zv_3)/2}
    \right]E_1.
    \label{eq:calx3}
\end{align}
In deriving the above expressions, we used Eq.~\eqref{eq:uveq0}.

Now, combining Eqs.~\eqref{eq:calx1}--\eqref{eq:calx3}, we obtain the equations of the form
\begin{align}
    \mathcal{X}_1+\left[\frac{1+NZ(v_1+v_2+Zv_3)/2}{1+NZv_2/2}\right]\mathcal{X}_2=
    (\cdots)E_1&=0,\label{eq:branch_condition_1}
    \\ 
    \mathcal{X}_1-\left[\frac{1+NZ(v_1+v_2+Zv_3)/2}{N(v_1-Zv_3)/2}\right]\mathcal{X}_3=
    (\cdots)E_1&=0.\label{eq:branch_condition_2}
\end{align}
Here, the ellipses denote messy expressions written in terms of $u_1,\dots,v_3$, and they do not vanish for generic $W_1$ and $W_2$.
Equations~\eqref{eq:branch_condition_1} and \eqref{eq:branch_condition_2} imply $E_1=0$ in a generic case in the space of 
$W_1$ and $W_2$. By themselves, 
they do not exclude all exceptional 
solutions with $E_1\neq0$.
As stated in the Introduction, 
this paper is restricted to the generic, 
regular branch and does not attempt an 
complete analysis of all exceptional cases.\footnote{The particular $E_1\neq0$ sector 
visible in the $W_1=W_2=0$ frame does 
not define an independent action on the branch 
for which the generalized disformal transformation to 
this frame is regular and invertible. 
Up to boundary terms, its contribution can be 
absorbed into redefinitions of coefficients already 
present in the action. Hence, $E_1=0$ may be imposed 
without loss of generality on this branch.}

Finally, setting $E_1=0$ and $e_2=2e_1'$, we have 
\begin{align}
    \mathcal{Y}&=
    \frac{2}{N}\left[1+\frac{NZ}{2}\left(v_1+v_2+Zv_3\right)\right]
    \frac{\partial\bar{\mathcal{G}}_3}{\partial Z}
    -\frac{1}{2}\left(3v_1+v_2+Zv_3\right)\bar{\mathcal{G}}_3
    \notag \\ & \quad 
    -\frac{1}{N}\frac{\partial(N\bar{\mathcal{G}}_4)}{\partial N}
    -2Z\left(u_1+Zu_2\right)\frac{\partial\bar{\mathcal{G}}_4}{\partial Z}
    +\left(u_1-Zu_2\right)\bar{\mathcal{G}}_4
    \notag \\ &=0,
    \label{eq:YG3G4}
\end{align}
where $\bar{\mathcal{G}}_3:=\mathcal{G}_3-Z\partial(Ne_1)/\partial N$
and $\bar{\mathcal{G}}_4:=\mathcal{G}_4-e_1$.
Thanks to the redundancy associated with adding the total divergence term~\eqref{eq:total-div} to the action,
one may set $e_1=0$ without loss of generality, which enforces $e_2=0$.

To find the associated potential for $\mathcal{G}_3$ and $\mathcal{G}_4$,
it is convenient to change a set of independent variables from $(N,Z)$ to $(N,Y)$, where $Y:=\ln Z-W_1(N,Z)$.
Equation~\eqref{eq:YG3G4} then reduces to
\begin{align}
    \frac{\partial}{\partial Y}
    \left(e^{-3W_1/2+W_2}\mathcal{G}_3\right)
    =\frac{1}{2}\frac{\partial}{\partial N}\left(
    NZe^{-3W_1/2+W_2}\mathcal{G}_4
    \right),
\end{align}
yielding
\begin{align}
    \mathcal{G}_3=e^{3W_1/2-W_2}\frac{\partial W_0(N,Y)}{\partial N},
    \quad 
    \mathcal{G}_4=\frac{2e^{3W_1/2-W_2}}{NZ}\frac{\partial W_0(N,Y)}{\partial Y},
\end{align}
where $W_0$ is an arbitrary function of $N$ and $Y$.
In terms of the original variables, we have
\begin{align}
    N\mathcal{G}_3&=e^{3W_1/2-W_2}
    \left[\frac{\partial W_0(N,Z)}{\partial \ln N}
    +\frac{\partial W_1/\partial \ln N}{1-\partial W_1/\partial\ln Z}
    \frac{\partial W_0(N,Z)}{\partial \ln Z}\right],
    \\
    NZ\mathcal{G}_4&=\frac{2e^{3W_1/2-W_2}}{1-\partial W_1/\partial\ln Z}
    \frac{\partial W_0(N,Z)}{\partial \ln Z}.
\end{align}

\subsection{Summary of the consequences of the consistency conditions}\label{subsec:summary}

Having described the technical calculations, we are now in a position to present the summary of the results obtained from the consistency conditions.
The Lagrangian satisfying the consistency conditions
(as well as the degeneracy conditions)
must be of the form
\begin{align}
    \mathcal{L}=\mathcal{K}^{ij}Q_{ij}+\frac{1}{2}\mathcal{K}^{ij,kl}Q_{ij}Q_{kl}
    +\mathcal{H}_1^{ij,kl}Q_{ij}a_{kl}
    +\mathcal{L}_0+\mathcal{L}_1, \label{eq:final_L}
\end{align}
where
\begin{align}
    Q_{ij}=K_{ij}+\mathcal{U}_{ij}V+\frac{1}{2}\mathcal{V}_{ij}^kV_k,
\end{align}
$\mathcal{L}_0$ contains neither $K_{ij}$, $V$, nor $V_i$,
and $\mathcal{L}_1$ is linearly dependent on $V$ and $V_i$:
\begin{align}
    \mathcal{L}_1=\frac{e^{3W_1/2-W_2}}{N}
    \left[\frac{\partial W_0(N,Z)}{\partial \ln N}
    +\frac{\partial W_1/\partial \ln N}{1-\partial W_1/\partial\ln Z}
    \frac{\partial W_0(N,Z)}{\partial \ln Z}\right]V
    +\frac{2}{NZ}
    \frac{e^{3W_1/2-W_2}}{1-\partial W_1/\partial\ln Z}
    \frac{\partial W_0(N,Z)}{\partial \ln Z}a^iV_i.
    \label{eq:Lag-1-lin}
\end{align}
Here, $\mathcal{K}^{ij}$, $\mathcal{K}^{ij,kl}$, and $\mathcal{H}_1^{ij,kl}$ are
any type I quantities written in the forms of Eqs.~\eqref{def:calKij},~\eqref{def:calKijkl}, and~\eqref{def:cal H1}, respectively,
while $\mathcal{U}_{ij}$ and $\mathcal{V}_{ij}^k$ must be the quantities of the forms~\eqref{eq:ansatz:Uij} and~\eqref{ansatz:cal Vijk}, respectively, with $u_1,\dots,v_3$ given by~\eqref{sol:u1}--\eqref{sol:u_1-v_3}.
We have three arbitrary functions $W_0$, $W_1$, and $W_2$, in addition to the 13 arbitrary functions of $N$ and $Z$ in $\mathcal{K}^{ij,kl}$, $\mathcal{K}^{ij}$, and $\mathcal{H}_1^{ij,kl}$.

In the next section, we see close relations between GD transformations and the functions $W_1$ and $W_2$.
More specifically, one can move to the frame in which $W_1=W_2=0$ by performing appropriate GD transformations.
We also demonstrate that integration by parts enables us to recast $\mathcal{L}_1$ into the form of the first term in Eq.~\eqref{eq:final_L}, and thus it is redundant, though this fact is not so evident at this moment.

\section{GD transformation properties and relation to GDH and U-DHOST theories}\label{sec:GD}

Let us discuss how the theories we have obtained in the previous section transform under GD transformations~\cite{Takahashi:2021ttd,Takahashi:2022mew}.

\subsection{GD transformation}

Disformal transformations of the metric~\cite{Bekenstein:1992pj, Bruneton:2007si, Bettoni:2013diz, Zumalacarregui:2013pma, Domenech:2015tca, BenAchour:2016cay, Takahashi:2017zgr} are translated into the transformations of the ADM variables as
\begin{align}
    N'=F_0(N),\qquad N_i'=F_1(N)N_i,\qquad \gamma_{ij}'=F_1\gamma_{ij}.
    \label{eq:dt-adm}
\end{align}
Generalized disformal transformations~\cite{Takahashi:2021ttd} are the following generalizations of the transformations~\eqref{eq:dt-adm}:
\begin{align}
        N'=F_0(N),\qquad N_i' = 
        F_1(N,Z)N_i+F_2(N,Z)a_i+F_3(N,Z)N^ja_ja_i,
        \qquad 
        \gamma_{ij}'=F_1\gamma_{ij}+F_3a_ia_j.
        \label{eq:gdt}
\end{align}
(In this section, primes are \textit{not} used to denote derivatives with respect to $Z$.)
Here, we restrict ourselves to the GD transformations that evade an extra Ostrogradsky mode in the presence of matter~\cite{Takahashi:2022mew,Takahashi:2022ctx,Naruko:2022vuh,Ikeda:2023ntu}.
We require that
\begin{align}
    F_1\neq 0,\qquad F_1+ZF_3\neq 0,\qquad 
    F_1-ZF_{1,Z}-Z^2F_{3,Z}\neq 0,\qquad  F_{0,N}\neq 0,
\end{align}
so that the inverse metric ${\gamma^{ij}}'$ exists and the transformation is invertible.

The transformation rules for various quantities under the transformations~\eqref{eq:gdt} are given by
\begin{align}
        {N^i}'&=N^i +\frac{F_2}{F_1+ZF_3}a^i,
        \\ 
        {\gamma^{ij}}'
        &=\frac{1}{F_1}\left(\gamma^{ij}-\frac{F_3}{F_1+ZF_3}a^ia^j\right),
        \\ 
        \sqrt{\gamma'}&=
        F_1\sqrt{F_1+ZF_3}\sqrt{\gamma},
        \\ 
        a_i'&=\nu_0a_i,
        \\ 
        a_{ij}'&=D_i(\nu_0a_j)-\nu_0 C_{ij}^ka_k,
        \\ 
        Z'&=\frac{\nu_0^2Z}{F_1+ZF_3},
        \\ 
        V'&=\nu_0 \left(V-\frac{ZF_2}{F_1+ZF_3}\right),
        \\ 
        V_i'&=D_i\left[ \nu_0\left(V-\frac{ZF_2}{F_1+ZF_3}\right)\right],
        \\
        K_{ij}'&=\frac{NF_1}{F_0}K_{ij}
        +\frac{F_3}{2F_0}\left(a_iV_j+a_jV_i\right)
        -\frac{1}{2F_0}\left[
                D_i(F_2a_j)+D_j(F_2a_i)-2F_2C_{ij}^ka_k
        \right]
        \notag \\ & \quad 
        +\frac{1}{2F_0}\left(
                F_1\nu_1\gamma_{ij}+F_3\nu_3a_ia_j
        \right)V
        +\frac{1}{ZF_0}\left(F_1\zeta_1\gamma_{ij}+F_3\zeta_3a_ia_j\right)
        \left(a^kV_k-Na^ka^lK_{kl}\right),
        \\ 
        {R^{(3)}}'&=\frac{1}{F_1}\left(
            \gamma^{ij}-\frac{F_3}{F_1+ZF_3}a^ia^j
        \right)\left(R_{ij}^{(3)}+D_kC^k_{ij}-D_{j}C_{ik}^k+C^k_{kl}C^l_{ij}-C^k_{il}C^l_{jk}\right),
        \label{transform:R}
\end{align}
where
\begin{align}
        C^{k}_{ij} 
        &=\frac{1}{2F_1}\left(\gamma^{kl}-\frac{F_3}{F_1+ZF_3}a^ka^l\right)
        \biggl[
                F_1\left(\nu_1a_i+\frac{2\zeta_1}{Z}a^ma_{im}\right)\gamma_{jl}
                +
                F_1\left(\nu_1a_j+\frac{2\zeta_1}{Z}a^ma_{jm}\right)\gamma_{il}
                \notag \\ & \quad 
                -F_1\left(\nu_1a_l+\frac{2\zeta_1}{Z}a^ma_{m}l\right)\gamma_{ij}
                +2F_3a_la_{ij}
                +F_3\left(\nu_3a_i+\frac{2\zeta_3}{Z}a^ma_{im}\right)a_ja_l
                \notag \\ & \quad 
                +F_3\left(\nu_3a_j+\frac{2\zeta_3}{Z}a^ma_{jm}\right)a_ia_l
                -F_3\left(\nu_3a_l+\frac{2\zeta_3}{Z}a^ma_{lm}\right)a_ia_j
        \biggr],
\end{align}
and 
\begin{align}
    \nu_n:=\frac{\partial \ln F_n}{\partial\ln N},
    \qquad 
    \zeta_n:=\frac{\partial\ln F_n}{\partial\ln Z}.
\end{align}

The above transformation rules indicate that the form of the unitary gauge Lagrangian~\eqref{eq:unitary-gauge-Lagrangian} is invariant under GD transformations.\footnote{As seen from Eq.~\eqref{transform:R},
a new term of the form $f(N,Z)a^ia^jR_{ij}$ is generated by GD transformations.
However, noting that $fa^ia^jR_{ij}=f(a^jD^ia_{ij}-a^iD_ia_j^j)$, one can recast this term into the terms that are already present in the Lagrangian by integration by parts.}
Tedious but straightforward calculations give us the transformation rules for the coefficient of each term in the Lagrangian.
Some of them are as simple as
\begin{align}
    \tilde b_1'&=\frac{F_0}{N}\frac{\tilde b_1}{F_1\sqrt{F_1+ZF_3}},
    \\
    \tilde b_2'&=\frac{F_0}{N}\frac{\tilde b_2}{F_1\sqrt{F_1+ZF_3}},
    \\ 
    b_4'&=\frac{F_0}{N}\frac{(F_1+ZF_3)^{3/2}}{F_1^2\nu_0^2}b_4
    +\frac{2F_0}{N}\frac{F_3\sqrt{F_1+ZF_3}}{F_1^2\nu_0^2}\tilde b_1,
    \\ 
    c_7'&=\frac{\sqrt{F_1+ZF_3}}{NF_1^2\nu_0^2}
    \left[NF_1c_7-F_3(2\tilde b_1+Zb_4)\right],
    \\ 
    h_1'&=\frac{\sqrt{F_1+ZF_3}}{F_1\nu_0}h_1
    +\frac{2F_2}{NF_1\sqrt{F_1+ZF_3}\nu_0}\tilde b_1,
    \\ 
    h_2'&=\frac{\sqrt{F_1+ZF_3}}{F_1\nu_0}h_2
    +\frac{2F_2}{NF_1\sqrt{F_1+ZF_3}\nu_0}\tilde b_2.
\end{align}
However, the transformation rules for the other coefficients are messy, and their explicit expressions are not illuminating.
One can verify that if one performs GD transformations to a theory satisfying the degeneracy and consistency conditions, then the resultant theory also satisfies the same conditions.
Thus, our general class of ghost-free scalar-tensor theories is stable under GD transformations.
One also sees that $\tilde b_2/\tilde b_1$ is invariant under GD transformations.
Furthermore, if $h_2/h_1=\tilde b_2/\tilde b_1$ in a frame, then the same relation holds in any frame.

The transformation rules for $W_0$, $W_1$, and $W_2$ can be derived by an explicit calculation.
For example, from
\begin{align}
    e^{W_2'}&=1+\frac{N'Z'}{2}v_2'=1+\frac{N'Z'c_7'}{2\tilde b_1'}+Z'b_4'
    \notag \\ & = \frac{F_1}{F_1+ZF_3}
    \left(
        1+\frac{NZc_7}{2\tilde b_1}+Zb_4
    \right)
    \notag \\ & = \frac{F_1}{F_1+ZF_3}
    e^{W_2},
\end{align}
we obtain the transformation rule
\begin{align}
    W_2'(N',Z')&=W_2(N,Z)
    -\ln\left[\frac{F_1(N,Z)+ZF_3(N,Z)}{F_1(N,Z)}\right].
\end{align}
Similarly, one finds that
\begin{align} 
    W_0'(N',Z')&=W_0(N,Z),
    \\
    W_1'(N',Z')&=W_1(N,Z)-\ln
    \left[F_1(N,Z)+ZF_3(N,Z)\right].
\end{align}
One can therefore move to the frame with $W_1=W_2=0$ by choosing $F_1$ and $F_3$ appropriately.
This result shows that all (vacuum) theories are equivalent to the subset with $W_1=W_2=0$ modulo GD transformations.

In the frame with $W_1=W_2=0$, it is easy to show that $\mathcal{L}_1$ can be removed by integration by parts.
When $W_1=W_2=0$, the Lagrangian~\eqref{eq:Lag-1-lin} reduces to
\begin{align}
    \mathcal{L}_1=\frac{\partial W_0}{\partial N}V
    +\frac{2}{N}\frac{\partial W_0}{\partial Z}a^iV_i,
\end{align}
which turns out to be equivalent to
\begin{align}
    -W_0K+2\frac{\partial W_0}{\partial Z}a^ia^jK_{ij},
    \label{eq:KaaK-lin1}
\end{align}
up to a total divergence.
This can be shown by noting that
\begin{align}
    \partial_t\left(\sqrt{\gamma}W_0\right)
    -\partial_i\left(N^i\sqrt{\gamma}W_0\right)
    &=
    \sqrt{\gamma}N
    \left[W_0K+\frac{\partial W_0}{\partial N}V+
    \frac{1}{N}
    \frac{\partial W_0}{\partial Z}
    \left(\partial_tZ-N^i\partial_iZ\right)\right]
    \notag \\ & 
    =\sqrt{\gamma}N\left[W_0K+
    \frac{\partial W_0}{\partial N}V+\frac{2}{N}\frac{\partial W_0}{\partial Z}
    \left(a^iV_i-Na^ia^jK_{ij}\right)
    \right].
\end{align}
Equation~\eqref{eq:KaaK-lin1} can then be absorbed into a redefinition of $g_1$ and $g_2$ in $\mathcal{K}^{ij}$.
Therefore, the terms in Eq.~\eqref{eq:Lag-1-lin} are in fact redundant.
Thus, after appropriate GD transformations and integration by parts, the Lagrangian can be recast into Eq.~\eqref{eq:final_L} with $\mathcal{L}_1=0$ and $Q_{ij}$ replaced by $K_{ij}$.

\subsection{The GDH action in the unitary gauge}

Let us present the action for the GDH theories constructed by applying a GD transformation to the Horndeski action~\cite{Takahashi:2022mew}.
The Horndeski action in the unitary gauge is of the form
\begin{align}
    S_{\textrm{H}}^{\textrm{(u.g.)}}[N,N_i,\gamma_{ij}]&=\int\D t\D^3x\sqrt{\gamma}N
    \left[
        g(N)K+b(N)\left(K^2-K_{ij}K^{ij}\right)+\mathcal{L}_0'
    \right],
    \\ 
    \mathcal{L}_0'&=f(N)R^{(3)}+k(N),
\end{align}
Here, we consider a subset of the Horndeski family of theories with $G_5=0$.
The Lagrangian $\mathcal{L}_0$ does not contain time derivatives, and hence is not important for the present discussion.
The Horndeski action is the particular case of our general action with
\begin{align}
    g_1'=g(N),\qquad \tilde b_1'=-\tilde b_2'=-b(N),
    \qquad 
    g_2'=g_3'=g_4'=\tilde b_3'=b_4'=b_5'=c_n'=h_n'=0.
\end{align}
This in particular implies that $W_1=W_2=0$ in the Horndeski theory.

The GDH action is then obtained from the Horndeski action as
\begin{align}
    S_{\textrm{GDH}}^{\textrm{(u.g.)}}[N,N_i,\gamma_{ij}]
    =S_{\textrm{H}}^{\textrm{(u.g.)}}[N',N_i',\gamma_{ij}'],
\end{align}
where $N'$, $N_i'$, and $\gamma_{ij}'$ are given by Eq.~\eqref{eq:gdt}.
Explicitly, the Lagrangian is given by
\begin{align}
    \mathcal{L}_{\textrm{GDH}}^{\textrm{(u.g.)}}&=
    \mathcal{K}^{ij}_{\textrm{GDH}}Q_{ij}+
    \frac{1}{2}\mathcal{K}^{ij,kl}_{\textrm{GDH}}Q_{ij}Q_{kl}
    +\mathcal{H}^{ij,kl}_{\textrm{GDH}}Q_{ij}a_{kl} + \mathcal{L}_0,\label{eq:GDH_L}
\end{align}
where $W_1$ and $W_2$ used to define $Q_{ij}$ are given by
\begin{align}
    W_1=\ln (F_1+ZF_3),\qquad W_2=\ln \left(\frac{F_1+ZF_3}{F_1}\right),
\end{align}
and the coefficients in $\mathcal{K}^{ij}_{\textrm{GDH}}$,
$\mathcal{K}^{ij,kl}_{\textrm{GDH}}$, and $\mathcal{H}^{ij,kl}_{\textrm{GDH}}$
are given by
\begin{align}
        &g_1=F_1\sqrt{F_1+ZF_3}g
        -\frac{ZF_1F_2\left[2F_1\nu_2+ZF_3(\nu_1+2\nu_2-\nu_3)\right]}{F_0(F_1+ZF_3)^{3/2}}b,
        \notag 
        \\ 
        &g_2=-\frac{F_1\left[3F_1\zeta_1+ZF_3(1+2\zeta_1+\zeta_3)\right]}{Z\sqrt{F_1+ZF_3}}g
        \notag \\ & \qquad 
        +\frac{F_1F_2}{F_0(F_1+ZF_3)^{3/2}}
        \left\{
                2F_1[(-1+\zeta_1)\nu_1+(1+2\zeta_1)\nu_2]+ZF_3
                [(1+2\zeta_1+2\zeta_3)\nu_1+(1+2\zeta_1)(2\nu_2-\nu_3)]
        \right\}b
        ,\notag 
\\
        &\tilde b_1=-\frac{N}{F_0}F_1\sqrt{F_1+ZF_3}b,
        \qquad 
        \tilde b_2=-\tilde b_1,
        \qquad 
        \tilde b_3=\frac{2\left[2F_1\zeta_1+ZF_3(1+\zeta_1+\zeta_3)\right]}{Z(F_1+ZF_3)}\tilde b_1,
        \notag \\ &
        b_4=-\frac{2F_3}{F_1+ZF_3}\tilde b_1,
        \qquad 
        b_5=-\frac{2\left\{3F_1\zeta_1^2+ZF_3[\zeta_3+\zeta_1(1+\zeta_1+2\zeta_3)]\right\}}{Z^2(F_1+ZF_3)}\tilde b_1,
        \notag 
\\
        &h_1=\frac{2F_1F_2}{F_0\sqrt{F_1+ZF_3}}b,
        \qquad 
        h_2=-h_1,
        \qquad 
        h_3=\frac{2F_1\zeta_1+ZF_3(1+\zeta_1+\zeta_3)}{Z(F_1+ZF_3)}h_1,
        \notag \\ &
        h_4=\frac{-2F_1\zeta_2+ZF_3(1-\zeta_1-2\zeta_2+\zeta_3)}{Z(F_1+ZF_3)}h_1,
        \qquad 
        h_5=\frac{2\zeta_1(\zeta_1+2\zeta_2)}{Z^2}h_1,
        \notag \\ & 
        h_6=-\frac{2\left[F_1(\zeta_1-\zeta_2)+ZF_3(1-\zeta_2+\zeta_3)\right]}{Z(F_1+ZF_3)}h_1.
\end{align}
Since these 13 functions can be arbitrary in the general class of theories we have constructed, it is clear that the GDH family represents only a subset of this class.

\subsection{U-DHOST theories}

Our general theories can be viewed as generalizations of U-DHOST theories~\cite{DeFelice:2018ewo, DeFelice:2021hps}.
The Lagrangian for U-DHOST theories is given by
\begin{align}
        \mathcal{L}^{(\textrm{u.g.})}_{\textrm{UD}}&=
        g_1(N)\gamma^{ij}Q_{ij}+\frac{1}{2}\mathcal{K}^{ij,kl}_{\textrm{UD}}
        Q_{ij}Q_{kl}+\mathcal{L}_0, \label{eq:L_U-DHOST}
        \\ 
        \mathcal{L}_0&=
        g_0(N)Z+f(N)R^{(3)}+k(N),
\end{align}
where 
\begin{align}
        \mathcal{K}^{ij,kl}_{\textrm{UD}}=\tilde b_1(N)\left(\gamma^{ik}\gamma^{jl}+\gamma^{il}\gamma^{jk}\right)
        +2\tilde b_2(N)\gamma^{ij}\gamma^{kl},
\end{align}
and
\begin{align}
        Q_{ij}=K_{ij}+\frac{1}{2}\frac{\partial W_1(N)}{\partial N}\gamma_{ij}V.
\end{align}
Note that $\mathcal{L}_0$ has nothing to do with the degeneracy and consistency conditions.
Note also that the term linearly dependent on $V$ is in fact redundant because it can be recast into the form $\tilde g_1(N)K$ by integration by parts.

Under disformal transformations~\eqref{eq:dt-adm}, the coefficients transform as
\begin{align}
    g_1\to g_1'=\frac{g_1}{F_1^{3/2}},
    \qquad 
    b_{1,2}\to b_{1,2}'=\frac{F_0}{NF_1^{3/2}}b_{1,2},
    \qquad 
    W_1\to W_1'=W_1-\ln F_1.
\end{align}
Furthermore, it is easy to verify that the form of $\mathcal{L}_0$ is invariant under disformal transformations thanks to the linear dependence on $Z$.
Therefore, U-DHOST theories are stable under disformal transformations.

Our general class of Lagrangians extends U-DHOST theories, with the characteristic field transformation generalized from disformal to GD transformations.
With this generalization, $\mathcal{K}^{ij}$ and $\mathcal{K}^{ij,kl}$ are allowed to depend on $a^i$, $Q_{ij}$ can depend on $a^i$ and linearly on $V_i$, and new terms of the form $\mathcal{H}_1^{ij,kl}Q_{ij}a_{kl}$ appear, resulting in a broader range of healthy interactions involving up to third-order derivatives of the scalar field.

\section{Discussion}\label{sec:discussion}

In this paper, we have systematically constructed
a class of ghost-free scalar-tensor theories 
whose Lagrangian contains up to third-order derivatives 
of the scalar field. Using a spatially 
covariant action in the unitary gauge, we imposed the 
degeneracy and consistency conditions~\cite{Motohashi:2014opa, Motohashi:2016ftl, Motohashi:2017eya, Motohashi:2018pxg,Gao:2018znj} needed to 
ensure the propagation of only one scalar and two tensor 
degrees of freedom, and determined the allowed form of the Lagrangian.

Each term in the resultant Lagrangian is expressed in terms of 
the extrinsic curvature $K_{ij}$ on constant-scalar-field hypersurfaces,
the ``velocity'' $V:=N^{-1}\left(\partial_tN-N^i\partial_iN\right)$ of 
the lapse function $N$, its spatial gradient $V_i:=D_iV$,
the acceleration $a_{i}:=D_i\ln N$, and its spatial derivative $a_{ij}:=D_ia_j$. 
After imposing all the conditions that remove the propagation of extra modes,
we found the general action with the following properties:
\begin{enumerate}[label=(\roman*)]
    \item Time derivatives of the spatial metric and the lapse function appear through a single combination of the form $Q_{ij}:=K_{ij}+\mathcal{U}_{ij}V+(1/2)\mathcal{V}_{ij}^k V_k$ in the action;
    \item The kinetic part of the general action is parameterized by 15 arbitrary functions $t$, $N$, and $Z:=a_ia^i$, 13 of which correspond to the coefficients of the terms linear and quadratic in $Q_{ij}$, and two of which completely determines the forms of $\mathcal{U}_{ij}$ and $\mathcal{V}_{ij}^k$;
    \item By performing appropriate generalized disformal (GD) transformations~\cite{Takahashi:2021ttd}, one can remove the latter two functions, making $\mathcal{U}_{ij}=\mathcal{V}_{ij}^k=0$.
\end{enumerate}
The third point does \textit{not} mean that it is sufficient to consider only a subset of theories with $\mathcal{U}_{ij}=\mathcal{V}_{ij}^k=0$, because GD transformations change the coupling to matter.
We have demonstrated how one can reproduce the two known families
of theories, generalized disformal Horndeski (GDH) theories~\cite{Takahashi:2022mew} and 
U-DHOST theories~\cite{DeFelice:2018ewo, DeFelice:2021hps}, 
by imposing the appropriate restrictions on the functions in the action.
Our construction thus provides a systematic 
framework for healthy scalar-tensor theories 
with third-order derivatives, extending the GDH and U-DHOST 
theories.

There are several directions for future developments. 
First, it would be crucial to study cosmological perturbations
in our general theories. We would then be able to explore 
the subclasses that are free from ghost and gradient instabilities in scalar and tensor 
perturbations, and in particular identify the subclasses in which the propagation 
speed of gravitational waves is equal to that of light.\footnote{On an FLRW background, 
$a_i=a_{ij}=V_i=0$ in the pure transverse-traceless 
tensor sector. Hence, terms containing $D_iV$ do not 
contribute to the quadratic tensor action.}
The former point is interesting, given that DHOST theories disformally disconnected from Horndeski exhibit instabilities of cosmological solutions~\cite{deRham:2016wji,Langlois:2017mxy},
while the latter point is phenomenologically important in light of
the simultaneous detection of the gravitational-wave event GW170817 and 
the gamma-ray burst 170817A emitted from a binary neutron star merger~\cite{LIGOScientific:2017vwq,LIGOScientific:2017ync,LIGOScientific:2017zic}. 
Second, it would be worthwhile to extend the vacuum theories treated 
in this paper to include coupling with matter fields. In general, 
when matter fields are coupled to degenerate theories, the degeneracy 
conditions are no longer satisfied in the matter sector, leading to 
the emergence of extra degrees of freedom~\cite{Deffayet:2020ypa,Garcia-Saenz:2021acj,Domenech:2025gao,Takahashi:2022mew,Takahashi:2022ctx,Naruko:2022vuh,Ikeda:2023ntu}. One should derive 
the conditions to avoid such a situation in our framework. 
Third, it would be important to investigate the Vainshtein screening mechanism~\cite{Vainshtein:1972sx,Babichev:2013usa}, which 
restores the predictions of general relativity in the vicinity of matter, in our theories.
In DHOST theories, 
the mechanism is broken inside astrophysical bodies~\cite{Kobayashi:2014ida, Koyama:2015oma,Saito:2015fza,Crisostomi:2017lbg,Langlois:2017dyl,Dima:2017pwp,Hirano:2019scf,Crisostomi:2019yfo}. 
It seems nontrivial how this mechanism works in our general theories.
Finally, it would be interesting to extend our theories to include up to 
third-order time derivatives of the scalar field, $\partial_t^3\phi$. 
In this paper, we have restricted the terms involving third-order derivatives of the scalar field to $V_{\mu}$ and $a_{\mu\nu}$ [Eq.~\eqref{eq:third_derivatives}] to avoid complications in the degeneracy conditions, 
while excluding terms containing third-order time derivatives of the 
scalar field ($\sim \pounds_nV$) and second-order time derivatives of the spatial metric ($\sim \pounds_nK_{\mu\nu}$). 
In contrast, in the context of point-particle mechanics, ghost-free theories with third-order time derivatives are constructed in Ref.~\cite{Motohashi:2017eya}. 
It might be possible to generalize this result to field theory, 
thereby constructing healthy theories with third-order time 
derivatives of the scalar field.

\acknowledgments
We thank Kazufumi Takahashi, Norihiro Tanahashi, and Zhibang Yao for fruitful discussions.
We also thank Hayato Motohashi for valuable comments.
The work of MM was supported by
the Rikkyo University Special Fund for Research.
The work of TK was supported by
JSPS KAKENHI Grant No.~JP25K07308 and
MEXT-JSPS Grant-in-Aid for Transformative Research Areas (A) ``Extreme Universe'',
No.~JP21H05182 and No.~JP21H05189.

\appendix

\section{Inverse of $\mathcal{K}^{ij,kl}$}\label{app:inverse-of-calK}

We show that the inverse of $\mathcal{K}^{ij,kl}$, i.e., $G_{ij,kl}$, does exist by
constructing it explicitly.
Let us assume that $G_{ij,kl}$ is of the form
\begin{align}
    G_{ij,kl}&=\frac{1}{2}
    \biggl[
    \varrho_1\left(\gamma_{ik}\gamma_{jl}+\gamma_{il}\gamma_{jk}\right)
                +\varrho_2\gamma_{ij}\gamma_{kl}
                +\varrho_3\left(\gamma_{ij}a_ka_l+\gamma_{kl}a_ia_j\right)
                \notag \\ & \quad 
               +\varrho_4\left(
                        \gamma_{ik}a_ja_l+\gamma_{jl}a_ia_k
                        +\gamma_{il}a_ja_k+\gamma_{jk}a_ia_l
                \right)+\varrho_5a_ia_ja_ka_l
    \biggr].
\end{align}
Then, we have
\begin{align}
    \mathcal{K}^{ij,mn}G_{mn,kl}&=\tilde b_1\varrho_1
    \left(\delta^i_k\delta^j_l+\delta^i_l\delta^j_k\right)
    +(\dots)\gamma^{ij}\gamma_{kl}
    +(\dots)\gamma^{ij}a_ka_l+(\dots)a^ia^j\gamma_{kl}
    \notag \\ &\quad 
    +(\dots)\left(\delta^i_ka^ja_l+\delta^i_la^ja_k+\delta^j_ka^ia_l+\delta^j_la^ia_k\right)
    +(\dots)a^ia^ja_ka_l
    \notag \\ & =
    \frac{1}{2}\left(\delta^i_k\delta^j_l+\delta^i_l\delta^j_k\right),
\end{align}
yielding \textit{six} equations relating $\{\tilde b_1,\tilde b_2,\tilde b_3,b_4,b_5\}$
with $\{\varrho_1,\varrho_2,\varrho_3,\varrho_4,\varrho_5\}$.
They are simultaneously satisfied by taking
\begin{align}
        \varrho_1&=\frac{1}{2\tilde b_1},
        \\ 
        \varrho_2&=\frac{1}{\tilde b_1\varrho_0}\left[
                -4\tilde b_1\tilde b_2-4\tilde b_2b_4Z+(\tilde b_3^2-4\tilde b_2b_5)Z^2
        \right],
        \\ 
        \varrho_3&=\frac{1}{\tilde b_1\varrho_0}\left[
                -2\tilde b_1\tilde b_3+4\tilde b_2b_4-(\tilde b_3^2-4\tilde b_2b_5)Z 
        \right],
        \\ 
        \varrho_4&=-\frac{b_4}{2\tilde b_1(2\tilde b_1+b_4Z)},
        \\ 
        \varrho_5&=\frac{1}{\tilde b_1\varrho_0 (2\tilde b_1+b_4Z )}
        \left\{
                4\tilde b_2b_4^2-8\tilde b_1^2b_5+\tilde b_1\left[
                        6\tilde b_3^2+8\tilde b_3b_4+4(b_4^2-6\tilde b_2b_5)
                \right]
                -b_4\left[\tilde b_3^2-4 (\tilde b_1+\tilde b_2 )b_5\right]Z
        \right\},
\end{align}
where
\begin{align}
    \varrho_0:=4\tilde b_1(\tilde b_1+3\tilde b_2)+
        \left[8\tilde b_2b_4+4\tilde b_1(\tilde b_3+b_4)\right]Z
        +\left[-2\tilde b_3^2+4(\tilde b_1+2\tilde b_2)b_5\right]Z^2.
\end{align}
The tensor $G_{ij,kl}$ can thus be constructed explicitly.
Note that we need to assume that
\begin{align}
    \tilde b_1\neq 0,\qquad 2\tilde b_1+b_4Z\neq 0,
    \qquad \varrho_0\neq 0.\label{eq:assumption_inverse}
\end{align}
These conditions are assumed implicitly throughout the paper.

\section{Variation of the action}\label{app:varDS}

The first variation of the action is given by
\begin{align}
    \frac{\delta S}{\delta K_{ij}}&=\sqrt{\gamma}N\left(
    \mathcal{K}^{ij}+\mathcal{K}^{ij,kl}K_{kl}+\mathcal{C}_5^{ij}V+\mathcal{C}_7^{ij,k}V_k
    +\mathcal{H}_1^{ij,kl}a_{kl}\right),
    \\ 
    \frac{\delta S}{\delta V}&=\sqrt{\gamma}N\left(
    g_3+2c_1V+c_4a^iV_i+\mathcal{C}_5^{ij}K_{ij}
    +\mathcal{H}_7^{ij}a_{ij}\right)
    \notag \\ & \quad 
    -\partial_i\left[\sqrt{\gamma}N
    \left(
        g_4a^i+c_4Va^i+2\mathcal{C}_2^{ij}V_j+\mathcal{C}_7^{kl,i}K_{kl}
        +\mathcal{H}_9^{kl,i}a_{kl}
    \right)\right].
\end{align}
Noting that $\mathcal{K}^{ij}, \mathcal{K}^{ij,kl}, \dots$ can be regarded as functions of $(\gamma_{ij},N,a_i)$,
it is convenient to define 
\begin{align}
    \mathcal{T}^{ij}_0&:=
        \frac{\partial(N\mathcal{K}^{ij})}{\partial N}
        +\frac{\partial(N\mathcal{K}^{ij,kl})}{\partial N}K_{kl}
        +\frac{\partial(N\mathcal{C}_5^{ij})}{\partial N}V
        +\frac{\partial(N\mathcal{C}_7^{ij,k})}{\partial N}V_k
        +N\frac{\partial \mathcal{H}_1^{ij,kl}}{\partial N}a_{kl},
        \\ 
    \mathcal{T}^i_0&:=
        \frac{\partial(Ng_4)}{\partial N}a^i 
        +\frac{\partial(Nc_4)}{\partial N}Va^i+
        2\frac{\partial(N\mathcal{C}_2^{ij})}{\partial N}V_j
        +\frac{\partial(N\mathcal{C}_7^{kl,i})}{\partial N}K_{kl}
        +N\frac{\partial \mathcal{H}_9^{kl,i}}{\partial N}a_{kl},
        \\
    \mathcal{T}^{ij,m}_1&:=
        \frac{\partial\mathcal{K}^{ij}}{\partial a_m}
        +\frac{\partial\mathcal{K}^{ij,kl}}{\partial a_m}K_{kl}
        +\frac{\partial\mathcal{C}_5^{ij}}{\partial a_m}V
        +\frac{\partial\mathcal{C}_7^{ij,k}}{\partial a_m}V_k
       +\frac{\partial\mathcal{H}_1^{ij,kl}}{\partial a_m}a_{kl}
        ,
        \\ 
    \mathcal{T}^{,i}_1&:=
        \frac{\partial g_3}{\partial a_i}+2\frac{\partial c_1}{\partial a_i}V
        +\left(\frac{\partial c_4}{\partial a_i}a^kV_k+c_4 V^i\right)
        +\frac{\partial \mathcal{C}_5^{kl}}{\partial a_i}K_{kl}
        {+\frac{\partial \mathcal{H}_7^{kl}}{\partial a_i}a_{kl}}
        ,
        \\
    \mathcal{T}^{i,j}_1&:=
        \frac{\partial g_4}{\partial a_j}a^i+g_4\gamma^{ij}
        +\frac{\partial c_4}{\partial a_j}Va^i+c_4V\gamma^{ij}
        +2\frac{\partial\mathcal{C}_2^{ik}}{\partial a_j}V_k
        +\frac{\partial\mathcal{C}_7^{kl,i}}{\partial a_j}K_{kl}
        {+\frac{\partial\mathcal{H}_9^{kl,i}}{\partial a_j}a_{kl}}
        ,
        \\
    \mathcal{T}^{ij,kl}_2&:=
    \frac{\partial\mathcal{K}^{ij}}{\partial\gamma_{kl}}
    +\frac{\partial\mathcal{K}^{ij,mn}}{\partial \gamma_{kl}}K_{mn}
    +\frac{\partial\mathcal{C}_5^{ij}}{\partial\gamma_{kl}}V
    +\frac{\partial\mathcal{C}_7^{ij,m}}{\partial \gamma_{kl}}V_m
    +\frac{\partial\mathcal{H}_1^{ij,mn}}{\partial \gamma_{kl}}a_{mn}
    \notag \\ &\quad +\frac{1}{2}\gamma^{kl}
    \left(
    \mathcal{K}^{ij}+\mathcal{K}^{ij,mn}K_{mn}+\mathcal{C}_5^{ij}V+\mathcal{C}_7^{ij,m}V_m
    +\mathcal{H}_1^{ij,mn}a_{mn}\right),
        \\
    \mathcal{T}^{,ij}_2&:=
        \frac{\partial g_3}{\partial \gamma_{ij}}+2\frac{\partial c_1}{\partial \gamma_{ij}}V
        +\frac{\partial c_4}{\partial \gamma_{ij}}a^kV_k
        {-}c_4a^{(i}V^{j)}+\frac{\partial\mathcal{C}_5^{kl}}{\partial\gamma_{ij}}K_{kl}
        +\frac{\partial\mathcal{H}_7^{kl}}{\partial\gamma_{ij}}a_{kl}
    \notag \\ &\quad 
        +\frac{1}{2}\gamma^{ij}\left(
        g_3+2c_1V+c_4a^kV_k+\mathcal{C}_5^{kl}K_{kl}
        +\mathcal{H}_7^{kl}a_{kl}\right),
        \\
    \mathcal{T}_2^{i,kl}&:=
        \frac{\partial g_4}{\partial \gamma_{kl}}a^i
        +\frac{\partial c_4}{\partial \gamma_{kl}}Va^i
        -g_4\gamma^{i(k}a^{l)}-c_4V\gamma^{i(k}a^{l)}
        +2\frac{\partial\mathcal{C}_2^{ij}}{\partial \gamma_{kl}}V_j+
        \frac{\partial \mathcal{C}_7^{mn,i}}{\partial \gamma_{kl}}K_{mn}
        +\frac{\partial \mathcal{H}_9^{mn,i}}{\partial \gamma_{kl}}a_{mn}
    \notag \\ & \quad 
    +\frac{1}{2}\gamma^{kl}\left(
        g_4 a^i+c_4Va^i+2\mathcal{C}_2^{ij}V_j+\mathcal{C}_7^{mn,i}K_{mn}
        +\mathcal{H}_9^{mn,i}a_{mn}
    \right).
\end{align}

The second variation of the action is obtained as
\begin{align}
    \frac{\delta^2S}{\delta K_{ij}(x)\delta V(y)}&=
    \sqrt{\gamma}N\mathcal{C}_5^{ij}(x)\delta(x-y)
    +\sqrt{\gamma}N\mathcal{C}_7^{ij,k}(x)\frac{\partial\delta(x-y)}{\partial x^k},
    \\ 
    \frac{\delta^2S}{\delta V(x)\delta V(y)}&=\sqrt{\gamma}\left[2N
    c_1-D_i\left(Nc_4a^i\right)\right]\delta(x-y)
    +\frac{\partial^2}{\partial x^i\partial y^j}\left[
       2 \sqrt{\gamma}N\mathcal{C}_2^{ij}\delta(x-y)
    \right].
\end{align}

To find the consistency conditions, we also need to calculate
$\delta^2S/\delta N(x)\delta K_{ij}(y)$,
$\delta^2S/\delta N(x)\delta V(y)$,
$\delta^2S/\delta \gamma_{ij}(x)\delta K_{kl}(y)$,
and 
$\delta^2S/\delta V(x)\delta\gamma_{ij}(y)$.
They are given by
\begin{align}
    \frac{\delta^2S}{\delta N(x)\delta K_{ij}(y)}&=
    \sqrt{\gamma}\mathcal{T}^{ij}_0\delta(x-y)
    -\sqrt{\gamma}a_m\mathcal{T}^{ij,m}_1\delta(x-y)
    +\sqrt{\gamma}\mathcal{H}_1^{ij,kl}a_ka_l\delta(x-y)
    \notag \\ & \quad 
    +\sqrt{\gamma}\mathcal{T}^{ij,m}_1(y)\frac{\partial\delta(x-y)}{\partial y^m}
    -2\sqrt{\gamma}\mathcal{H}_1^{ij,kl}a_k(y)\frac{\partial\delta(x-y)}{\partial y^l}
    \notag \\ & \quad 
    -\sqrt{\gamma}\left[
        \frac{\partial\mathcal{H}_1^{ij,km}}{\partial y^k}
        +\Gamma_{kl}^l\mathcal{H}_1^{ij,km}+\Gamma_{kl}^m\mathcal{H}_1^{ij,kl}
    \right](y)\frac{\partial\delta(x-y)}{\partial y^m}
    -\frac{\partial^2}{\partial x^k\partial y^l}\left[
        \sqrt{\gamma}\mathcal{H}_1^{ij,kl}\delta(x-y)
    \right],
    \\ 
    \frac{\delta^2S}{\delta N(x)\delta V(y)}&=\left(\dots\right)\delta(x-y)+
    \sqrt{\gamma}\mathcal{T}^{,i}_1(y)\frac{\partial\delta(x-y)}{\partial y^i}
    - \sqrt{\gamma}\mathcal{T}^i_0(y)
    \frac{\partial\delta(x-y)}{\partial y^i}
    +\sqrt{\gamma}\mathcal{T}^{i,j}_1 a_j(y)\frac{\partial\delta(x-y)}{\partial y^i}
    \notag \\ & \quad 
    -2\sqrt{\gamma}\mathcal{H}_7^{ij}a_j(y)\frac{\partial\delta(x-y)}{\partial y^i}
    -\sqrt{\gamma}D_j\mathcal{H}_7^{ij}(y)\frac{\partial\delta(x-y)}{\partial y^i}
    -\sqrt{\gamma}\mathcal{H}_9^{kl,i}a_ka_l\frac{\partial\delta(x-y)}{\partial y^i}
    \notag \\ & \quad 
    +
    \frac{\partial^2}{\partial y^i\partial x^j}\left[
        \sqrt{\gamma}
        \left(\mathcal{T}^{i,j}_1-\mathcal{H}_{7}^{ij}-2\mathcal{H}_9^{kj,i}a_k
        -\Gamma_{kl}^j\mathcal{H}_9^{kl,i}
        \right)\delta(x-y)
    \right]-\frac{\partial^2}{\partial x^j\partial x^k}
        \frac{\partial}{\partial y^i}\left[\sqrt{\gamma}\mathcal{H}_9^{jk,i}\delta(x-y)\right],
    \\ 
    \frac{\delta^2S}{\delta\gamma_{ij}(x)\delta K_{kl}(y)}
    &=\sqrt{\gamma}N\left[\mathcal{T}_2^{kl,ij}
    +\mathcal{H}_1^{kl,mn}a^{(i}\Gamma_{mn}^{j)}
    \right]\delta(x-y)
    -\frac{1}{2}\sqrt{\gamma}N\left[
        2\mathcal{H}_1^{kl,m(i}a^{j)}-\mathcal{H}_1^{kl,ij}a^m
    \right](y)\frac{\partial\delta(x-y)}{\partial y^m}
    ,
    \\ 
    \frac{\delta^2S}{\delta V(x)\delta\gamma_{ij}(y)}
    &=\sqrt{\gamma}N\mathcal{T}_2^{,ij}\delta(x-y)+\sqrt{\gamma}N\mathcal{T}_2^{m,ij}(y)
    \frac{\partial\delta(x-y)}{\partial y^m}
    +\frac{1}{2}\sqrt{\gamma}N\left[
    2\mathcal{H}_7^{m(i}a^{j)}-a^m\mathcal{H}_7^{ij}
    \right](x)\frac{\partial\delta(x-y)}{\partial y^m}
    \notag \\ & \quad 
    +
    \sqrt{\gamma}N\mathcal{H}_9^{kl,m}a^{(i}\Gamma_{kl}^{j)}(y)
    \frac{\partial\delta(x-y)}{\partial y^m}
    -\frac{1}{2}\frac{\partial^2}{\partial x^l \partial y^k}
    \left[
    \sqrt{\gamma}N\left(
    2a^{(i}\mathcal{H}_9^{j)k,l}-a^k\mathcal{H}_9^{ij,l}
    \right)\delta(x-y)
    \right]
    .
\end{align}
In $\delta^2S/\delta N(x)\delta V(y)$, we do not present the terms
that are proportional to $\delta(x-y)$, because they do not contribute to $\mathcal{F}(x,y)$.

\section{Derivation of the solution to $\mathcal{A}^i_{kl}=\mathcal{B}^{ij}_{kl}=0$}\label{app:der1}

First, notice that
$N\times\,$Eq.~\eqref{eq:uveq1}
$+N\times\,$Eq.~\eqref{eq:uveq2}
$+NZ\times\,$Eq.~\eqref{eq:uveq3}
gives 
\begin{align}
    &\left[N\left(u_1+NZu_2\right)\right]'\left[1+\frac{NZ}{2}\left(v_1+v_2+Zv_3\right)\right]
    -\left[N\left(u_1+NZu_2\right)\right]\left[1+\frac{NZ}{2}\left(v_1+v_2+Zv_3\right)\right]'
    \notag \\ &
    =\frac{1}{2Z}\frac{\partial}{\partial \ln N}
    \left[1+\frac{NZ}{2}\left(v_1+v_2+Zv_3\right)\right],
\end{align}
which can be written as 
\begin{align}
    \frac{\partial}{\partial\ln Z}
    \left[
        \frac{N\left(u_1+NZu_2\right)}{1+NZ\left(v_1+v_2+Zv_3\right)/2}
    \right]
    =-\frac{1}{2}\frac{\partial}{\partial \ln N}
    \left[
        \frac{1}{1+NZ\left(v_1+v_2+Zv_3\right)/2}
    \right].
    \label{eq:app-ints1}
\end{align}
Let us define a function $W_1(N,Z)$ by
\begin{align}
    \frac{\partial W_1}{\partial \ln Z}
    :=\frac{NZ\left(v_1+v_2+Zv_3\right)}{2+NZ\left(v_1+v_2+Zv_3\right)}.
\end{align}
Then, we have
\begin{align}
    NZ\left(v_1+v_2+Zv_3\right)=\frac{2\partial W_1/\partial\ln Z}{1-\partial W_1/\partial\ln Z}.
    \label{eq:app-soln1}
\end{align}
Substituting this into Eq.~\eqref{eq:app-ints1}, we obtain 
\begin{align}
    \frac{\partial}{\partial \ln Z}
    \left[
    N(u_1+Zu_2)(1-\partial W_1/\partial\ln Z)
    \right]
    =\frac{1}{2}\frac{\partial^2W_1}{\partial\ln Z\partial \ln N}.
\end{align}
This can be integrated to give 
\begin{align}
    N(u_1+Zu_2)=\frac{1}{2}\frac{\partial W_1/\partial\ln N+w_1(N)}{1-\partial W_1/\partial\ln Z}.
    \label{eq:app-soln2}
\end{align}
Here, $w_1(N)$ can be absorbed into a redefinition of $W_1$.
From Eqs.~\eqref{eq:uveq0} and~\eqref{eq:app-soln1}, we obtain
\begin{align}
    \frac{NZ}{2}v_1&=\frac{\partial W_1/\partial \ln Z-\partial W_2/\partial\ln Z}{1-\partial W_1/\partial \ln Z},
    \\ 
    \frac{NZ^2}{2}v_3&=1-e^{W_2}+
    \frac{\partial W_2/\partial\ln Z}{1-\partial W_1/\partial \ln Z},
\end{align}
where we defined a function $W_2$ by
\begin{align}
     \frac{NZ}{2}v_2&=e^{W_2}-1.
\end{align}
From Eqs.~\eqref{eq:uveq2} and~\eqref{eq:app-soln2}, we obtain
\begin{align}
    2Nu_1&=\left(\frac{\partial W_1/\partial\ln N}{1-\partial W_1/\partial\ln Z}
    \right)
    \left(
        1-\frac{\partial W_2}{\partial\ln Z}
    \right)-\frac{\partial W_2}{\partial \ln N},
    \\
    2NZu_2&=\left(\frac{\partial W_1/\partial\ln N}{1-\partial W_1/\partial\ln Z}\right)
    \frac{\partial W_2}{\partial\ln Z}
    +\frac{\partial W_2}{\partial \ln N}. 
\end{align}
Now one can see that Eqs.~\eqref{eq:uveq1} and~\eqref{eq:uveq3} are satisfied.

In the above, $NZ\left(v_1+v_2+Zv_3\right)\neq -2$ is assumed.
Let us consider the special case with $NZ\left(v_1+v_2+Zv_3\right)=-2$.
In this case, Eq.~\eqref{eq:uveq0} implies that
\begin{align}
    (NZv_1+2)(NZv_2+2)=0,\label{eq:special_case}
\end{align}
and therefore we have the following two cases:
(i) $v_1=-2/NZ$ and $v_2+Zv_3=0$, and
(ii) $v_2=-2/NZ$ and $v_1+Zv_3=0$.
In both cases, Eq.~\eqref{eq:uveq3} amounts to a relation between $u_1$ and $u_2$.
Thus, we have two free functions, $v_3$ and $u_2$, in the special case with
$NZ\left(v_1+v_2+Zv_3\right)= -2$.

The exceptional branch satisfying $NZ(v_1+v_2+Zv_3)=-2$ is not included in the regular class considered here because it has no smooth homogeneous limit. In a local orthonormal frame with $a_i=\sqrt Z \delta_i^1$, Eq.~\eqref{eq:special_case} yields either 
$\mathcal V^1_{AB}=-2\delta_{AB}/(N\sqrt Z)$ or $\mathcal V^A_{1A}=-1/(N\sqrt Z)$. Thus, $\mathcal V^k_{ij}$ diverges as $Z^{-1/2}$ when $a_i\to0$.
We do not claim that this exceptional branch is inconsistent. Its remaining consistency conditions require a separate analysis, and a complete classification of such singular branches is outside the scope of the present work.

\bibliography{refs}
\bibliographystyle{JHEP}

\end{document}